\documentclass[sigconf,screen,nonacm]{acmart}

\usepackage{dirtree}
\usepackage{booktabs}
\usepackage{siunitx}
\usepackage{graphicx}  
\usepackage{enumitem}
\usepackage{subcaption}
\usepackage{multirow}
\usepackage{tabularx} 
\usepackage{array}

\newcommand{\name}{\texttt{EyeNavGS}}
\newcommand{\Lname}{\texttt{EyeNavGS}}

\newcommand{\threemeterarea}{3\,m $\times$ 3\,m\ }

\AtBeginDocument{%
	}

\begin{document}
	\sloppy
	
	\title{\texttt{EyeNavGS}: A 6-DoF Navigation Dataset and Record-n-Replay Software for Real-World 3DGS Scenes in VR}
	
	\settopmatter{authorsperrow=3}
	
	\author{Zihao Ding}
	\affiliation{
		\institution{Rutgers University}
		\city{Piscataway} 
		\state{NJ}      
		\country{USA}
	}
	
	\author{Cheng-Tse Lee}
	\affiliation{
		\institution{National Tsing Hua University}
		\city{Hsinchu}
		\country{Taiwan}
	}
	
	\author{Mufeng Zhu}
	\affiliation{
		\institution{Rutgers University}
		\city{Piscataway}
		\state{NJ}
		\country{USA}
	}
	
	\author{Tao Guan}
	\affiliation{
		\institution{Rutgers University}
		\city{Piscataway}
		\state{NJ}
		\country{USA}
	}
	
	\author{Yuan-Chun Sun}
	\affiliation{
		\institution{National Tsing Hua University}
		\city{Hsinchu}
		\country{Taiwan}
	}
	
	\author{Cheng-Hsin Hsu}
	\affiliation{
		\institution{National Tsing Hua University}
		\city{Hsinchu}
		\country{Taiwan}
	}
	
	\author{Yao Liu}
	\affiliation{
		\institution{Rutgers University}
		\city{Piscataway}
		\state{NJ}
		\country{USA}
	}
	
	\renewcommand{\shortauthors}{Ding et al.}
	
	\begin{abstract}
		3D Gaussian Splatting (3DGS) is an emerging media representation that reconstructs real-world 3D scenes in high fidelity, enabling 6-degrees-of-freedom (6-DoF) navigation in virtual reality (VR).
		However, developing and evaluating 3DGS-enabled applications and optimizing their rendering performance, require realistic user navigation data. Such data is currently unavailable for photorealistic 3DGS reconstructions of real-world scenes.
		This paper introduces \name, the first publicly available 6-DoF navigation dataset featuring traces from 46 participants exploring twelve diverse, real-world 3DGS scenes. 
		The dataset was collected at two sites, using the Meta Quest Pro headsets, recording the head pose and eye gaze data for each rendered frame during free world standing 6-DoF navigation. 
		For each of the twelve scenes, we performed careful scene initialization to correct for scene tilt and scale, ensuring a perceptually-comfortable VR experience. 
		We also release our open-source SIBR viewer software fork with record-and-replay functionalities and a suite of utility tools for data processing, conversion, and visualization. 
		The \name\  dataset and its accompanying software tools provide valuable resources for advancing research in 6-DoF viewport prediction, adaptive streaming, 3D saliency, and foveated rendering for 3DGS scenes. The \name\ dataset is available at: \textbf{\url{https://symmru.github.io/EyeNavGS/}}.
	\end{abstract}
	
	\maketitle
	
	\section{Introduction}
	
	Since its introduction in 2023, 3D Gaussian Splatting (3DGS)~\cite{kerbl20233d} has quickly emerged as a popular immersive media format for 3D scene representation, enabling high-fidelity, 6-degrees-of-freedom (6-DoF) exploration of complex real-world environments. 
	Due to its fast training time and real-time rendering speed, it has received significant attention from both academia and industry~\cite{MarvieGautron2025GaussianSplatting,404GenTextTo3D,kocabas2024hugs,yu2024mip,huang20242d,kerbl2024hierarchical,fang2024mini}. 
	
	3DGS has unlocked new possibilities, including rendering on mobile devices with WebGL support~\cite{antimatter15Splat,mkkelloggGaussianSplats3D,gsplatTech,papantonakis2024reducing} and extending traditional video streaming paradigms to full 6-DoF volumetric content. 
	For example, recent works such as SGSS~\cite{zhu2025sgss} and L3GS~\cite{tsai2025l3gs} have proposed streaming approaches for static 3DGS scenes. LapisGS~\cite{shi2024lapisgs} introduced a layered 3DGS representation that supports progressive adaptive streaming. Building on LapisGS, LTS~\cite{sun2025lts} proposed approaches for adaptive streaming of dynamic 3DGS scenes.
	
	However, the development and evaluation of these 3DGS-enabled systems and applications are hampered by the lack of suitable datasets. 
	To properly assess system performance of adaptive streaming algorithms, rendering optimizations, compression strategies, and quality of experience under real-world conditions, large-scale datasets recording authentic user interaction with  6-DoF scenes are essential. 
	To the best of our knowledge, no publicly available dataset currently captures such 6-DoF user navigation traces for real-world scenes reconstructed by 3DGS.
	The absence of such datasets forces researchers to rely on synthetic traces~\cite{lin2025metasapiens,zhu2025sgss} or datasets collected from different 3D representations~\cite{KhanChakareski2020NJIT6DOF}, which may not faithfully represent user interactions with high-fidelity 3DGS content.
	
	To close this gap, this paper introduces \name, the first publicly available dataset of user navigation traces. The dataset includes traces through twelve scenes. These scenes include both indoor and outdoor environments, offering diverse visual characteristics for studying user navigation behaviors and performance-quality tradeoffs in virtual reality (VR). Our contributions are summarized as follows:
	\vspace{1em}
	\begin{itemize}[leftmargin=*,topsep=0pt]
		\item \textbf{The \name\ Dataset.} We collected navigation traces of \textbf{46} participants. Traces were collected at two physical locations. Each trace includes a human user's exploration of twelve diverse indoor and outdoor scenes reconstructed by 3DGS. Each scene underwent careful initialization for tilt correction, metric scale establishment, and starting viewpoint selection to ensure perceptually comfortable VR experiences. The dataset includes per-frame head pose and eye gaze data, captured with Meta Quest Pro headsets during free world standing exploration.
		\item \textbf{The \name\ Record-n-Replay Software.} We release our open-source software, a fork of the SIBR viewer~\cite{sibr2020} for 3DGS, enhanced with record-and-replay functionality. 
		This fork includes capabilities for recording user traces and replaying these traces frame-by-frame for visualization, video generation, and detailed offline analysis.
		\item \textbf{The \name\ Utility Tools.} In addition to the core software, we provide a suite of utility tools. These tools include conversion operations to ease integration with other frameworks and allow reproduction and visualization of collected traces.
	\end{itemize}
	
	We anticipate this dataset, along with the accompanying software and tools, will facilitate more reliable and comprehensive evaluations of 6-DoF viewport prediction, view-adaptive streaming, 3D saliency, and foveated rendering for 3DGS.  
	We also encourage collaborative expansion of this dataset, aiming to create a richer collection of data for advancing research in immersive media experiences. 
	
	\section{Related Work}
	\label{sec:related}
	\subsection{6-DoF Navigation Datasets}
	
	The importance of 6-DoF navigation datasets in evaluating the streaming performance and user experience in immersive environments is well-recognized in the research community in recent years. While several 6-DoF navigation datasets have been created to date, each of them has their distinct focus and limitations.
	
	\vspace{0.3em}
	\noindent{\textbf{User navigation with synthetically generated environments.}}
	Khan and Chakareski~\cite{KhanChakareski2020NJIT6DOF,chakareski20206dof} introduced the ``NJIT 6DOF VR Navigation Dataset'', which recorded 6-DoF traces of three users exploring a synthetic ``Virtual Museum'' (sourced from the Unity Asset Store) using an HTC Vive wireless VR headset. 
	Similarly, Chen et al.~\cite{Chen22VRViewportPose} collected the VRViewportPose dataset, recording viewing traces of 30 participants on three different platforms, a desktop, an Oculus Quest 2 VR headset, and an Android smartphone as they interacted with three VR games with synthetic scenes. 
	Most recently, Ouellette et al.~\cite{ouellette2025mazelab} created a point cloud video dataset with user behavior traces collected via a Meta Quest 2 headset. The environment in this dataset mainly consists of a synthetically modeled maze.
	
	While these datasets, focusing on synthetic scenes, offer valuable insights into user navigation behaviors, they do not capture interactions within reconstructed representations of real-world scenes. Real-world characteristics, such as fine-grained textural details or subtle lighting variations, can influence user behavior~\cite{lessels2004changes}. Such differing user behaviors could cause researchers to draw misleading conclusions about the real-world effectiveness of 3DGS systems. 
	
	\vspace{0.3em}
	\noindent{\textbf{User interaction with dynamic point clouds.}}
	Subramanyam et al.~\cite{subramanyam2020user} created a 6-DoF navigation dataset where users viewed 150 frames of dynamic point cloud sequences from the 8i dataset~\cite{8i} using an Oculus Rift headset. 
	The 8i dataset features point cloud representations of individual human subjects. As a result, user navigation typically follows an ``outside-looking-in'' pattern. Thus, the viewing patterns are likely to be substantially different from  free exploration of expansive virtual worlds.
	
	Hu et al. released volumetric video viewing behavior dataset~\cite{hu2023understanding}
	, recorded with the Meta Quest Pro headset. 
	In this dataset, 50 participants watched 26 volumetric videos, represented as point clouds, from the FSVVD dataset~\cite{hu2023fsvvd}. A limitation 
	of using point clouds to collect these navigation datasets lies in their
	their relatively low rendered visual quality. Even with careful calibration and alignment, the produced point clouds cannot render views at a photorealistic quality comparable to emerging immersive media representations, such as neural radiance fields (NeRF)~\cite{mildenhall2020nerf} and 3DGS.
	
	In summary, while several 6-DoF user navigation datasets have been collected using synthetic environments and dynamic point clouds, they do not address navigation within photorealistic reconstructions of real-world scenes.
	Our dataset bridges this gap by providing recorded traces of users navigating 3DGS scenes.

	\subsection{OpenXR and SIBR Viewer for 3DGS in VR}
	
	\begin{figure}[!t]
		\centering
		\includegraphics[width=0.45\textwidth]{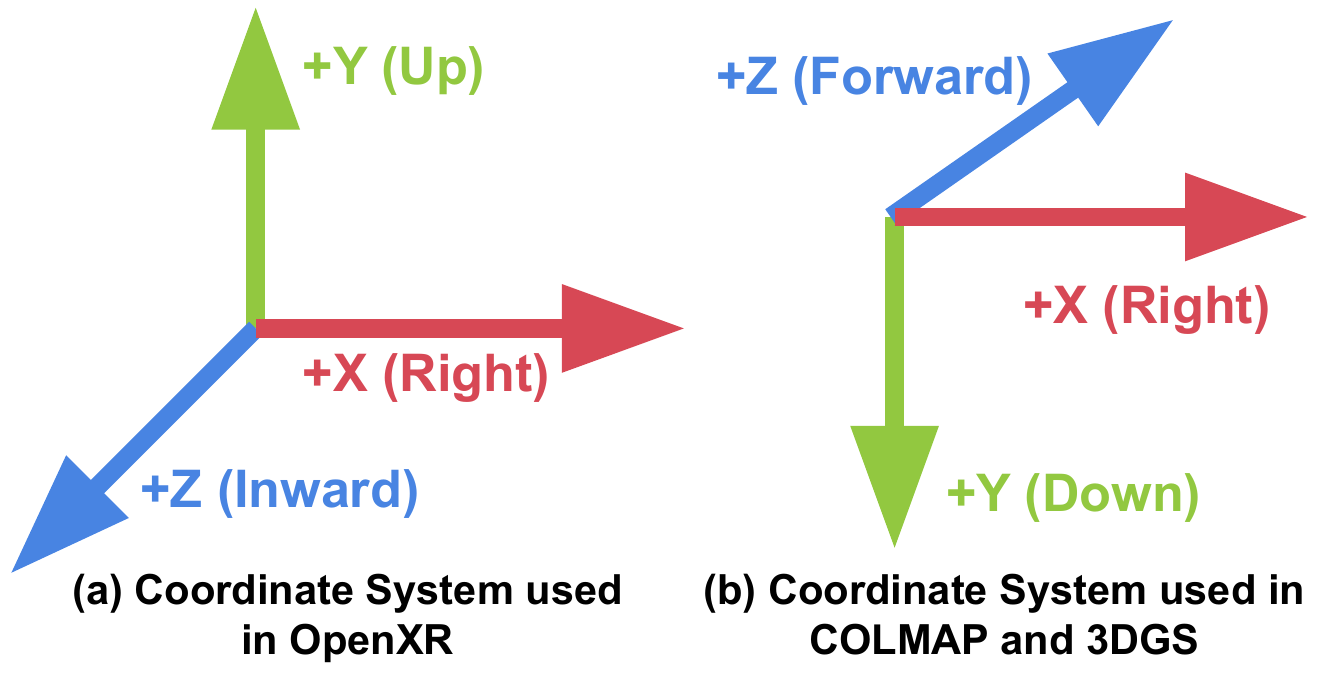}
		\vspace{-1em}
		\caption{The SIBR viewer source code resolves the mismatch between OpenXR and 3DGS's coordinate systems by rotating the camera 180 degrees around the x-axis.}
		\vspace{-1em}
		\label{fig:coordinate-systems}
	\end{figure}
	
	To experience 3DGS scenes immersively in VR, we use the open-source SIBR viewer~\cite{sibr2020}. 
	Specifically, our data collection software is a fork of its \texttt{gaussian\_code\_release\_openxr} branch, which renders 3DGS views to head-mounted displays (HMDs) via OpenXR~\cite{KhronosOpenXRPage}. 
	OpenXR provides a standardized API for VR and augmented reality (AR) applications and defines several reference coordinate spaces. 
	Figure \ref{fig:coordinate-systems}(a) shows the OpenXR coordinate systems, which are right-handed. 
	In particular, OpenXR defines three main types of reference spaces:
	
	\begin{itemize}[leftmargin=*]
		\item\textbf{View Space.}
		The space is relative to the user's head. 
		For stereo VR headsets, its origin is centered between two eyes.
		The axes are defined as +X to the right, +Y up, and -Z in the forward viewing direction. 
		
		\item\textbf{Local Space.}
		For VR devices that support 3-DoF rotational tracking only, they only support the ``Local Space'', where the headset is locked to a fixed origin in the world, typically the user's starting position, with the +Y axis aligned with gravity. 
		It is suitable for stationary or ``seated'' experiences where the user does not physically walk around. 
		
		\item\textbf{Stage Space.}
		For VR devices that support the full 6-DoF tracking, they can support the ``Stage Space''. The ``Stage Space'' defines a flat, rectangular area on the physical floor that the user can freely walk within--analogous to a performance stage.
		The XZ plane is aligned with the floor, the +Y axis defines the ``up'' direction, and the origin is fixed relative to the physical space.
		The ``Stage Space'' allows an application to  use tracked physical movements (position and orientation).
	\end{itemize}
	
	\noindent\textbf{SIBR VR Viewing Modes.}
	Using the ``Local'' and ``Stage'' spaces, the SIBR \texttt{gaussian\_code\_release\_openxr} branch supports two VR viewing modes: \texttt{seated} and \texttt{free world standing} (\texttt{fws}). 
	The \texttt{seated} mode is for stationary use, e.g., for VR headsets that supports 3-DoF (i.e., rotational) tracking only, mapping head rotation to orientation and controller input to position. 
	
	The \texttt{free world standing}  (\texttt{fws}) mode uses the ``Stage Space'' to map the user's tracked physical movements 1:1 to the virtual scene and requires a VR headset that supports the full 6-DoF (i.e., both rotational and positional) tracking. 
	
	For our data collection, we used the Meta Quest Pro, a headset with 6-DoF tracking capabilities. We thus only used the \texttt{fws} mode to capture users' natural physical movements in our trace collection. 
	
	\vspace{0.3em}
	\noindent\textbf{Coordinate Systems Mismatch.}
	Gaussians in a 3DGS scene are trained from an initial point cloud obtained via COLMAP~\cite{schoenberger2016mvs,schoenberger2016sfm}. Thus, 3DGS inherits COLMAP's coordinate system, which is a right-handed  system where the +Y axis points downwards and the +Z axis points forward (as shown in Figure \ref{fig:coordinate-systems}(b)). 
	This convention conflicts with OpenXR's coordinate systems, where +Y points upwards. 
	To resolve this mismatch, 
	the SIBR viewer source code rotates the camera 180 degrees around the X-axis to ensure the scene is rendered upright to the user.

	\section{\Lname\ Record-n-Replay Software}
	
	\label{sec:software}
	\subsection{Scene Initialization}
	\label{sec:initialization}
	To prepare each trained 3DGS scene for immersive exploration, we correct each raw scene to align with human assumptions about the physical world. These corrections include i) correcting the initial quaternion to fix scene tilt, ii) selecting a per-scene scale factor to ensure objects match their real-world proportions, and iii) establishing an example starting viewing position.
	Values for each scene are shown in Table \ref{tbl:scene_initialization}. 
	
	\vspace{0.3em}
	\noindent\textbf{Scene Tilt and Orientation Correction.}
	3DGS scenes are trained from initial point clouds generated by COLMAP~\cite{schoenberger2016sfm,schoenberger2016mvs}. 
	However, the coordinate system reconstructed by COLMAP is not inherently gravity-aligned. 
	Improperly oriented scenes frequently result in disorienting tilts, unnatural slopes, skewed camera behavior that degrade the sense of presence and spatial coherence within the virtual environment.
	To create a perceptually comfortable experience in VR, 
	we must first align the virtual scene with gravity. 
	
	Instead of modifying the trained 3DGS \texttt{.ply} file for each scene, our solution is to apply a corrective transformation when each scene is loaded. 
	More specifically, the initial quaternion, which rectifies any scene tilt and establishes a level ground plane, is set via the \texttt{poseInReferenceSpace} member of the \texttt{XrReferenceSpaceCreateInfo} structure.
	
	To find the amount of scene tilt, we implemented a robust procedure using Blender~\cite{BlenderWebsite} and the KIRI Engine add-on~\cite{KiriEngine3DGS}, which supports 3DGS point data.
	Within Blender, we inserted a reference plane perpendicular to the Y axis (which aligns with gravity) and manually adjusted the scene's orientation to ensure its ground plane is orthogonal to the virtual Y-axis and matches the reference plane. 
	This process effectively corrected any residual tilt, ensuring that users perceive the scene as grounded and stable, avoiding perceptual illusions of being on a slope. 
	These rotation parameters were subsequently exported as quaternions and applied at runtime during VR rendering to ensure proper alignment with the user's physical ``stage'' area.

	\begin{table}[!t]
		\centering
		\caption{Scene Initialization Parameters}
		\label{tbl:scene_initialization}
		\vspace{-0.5em}
		\resizebox{0.48\textwidth}{!}{%
			\begin{tabular}{cccc}
				\toprule
				\multirow{2}{*}{\textbf{Scene}} & \textbf{Initial Quaternion} & \multirow{2}{*}{\textbf{Scale}} & \textbf{Example Init. Pos.} \\
				& $(q_x, q_y, q_z, q_w)$ & & $(x,y,z)$ \\
				\midrule
				truck & -0.0896, 0, 0, 0.9960 & 0.76 &  0, 2.1, -4  \\
				treehill & -0.1961, 0, 0, 0.9806 & 12 & 2, 1.4, 2 \\
				train & 0.0499, 0, 0.01, 0.9987 & 0.36 & 2, -1, 6\\
				stump & -0.3950, 0, 0, 0.9187 & 1 & -1, 2.65, -2.5 \\
				room & -0.2334, 0, 0, 0.9724 & 2 & 0, 1.15, 0 \\
				playroom & -0.1961, 0, 0, 0.9806 & 2.7 & 0, 0.88, 0 \\
				drjohnson & -0.3699, 0, 0.5976, 0.7114 & 1 & 0, 1.5, 0\\
				bicycle & -0.1142, 0, 0, 0.9935 & 1.25 & 1.5, 1.1, 0\\
				nyc & -0.1483, 0, 0, 0.9888 & 0.64 & -1.6, 4.4, 4 \\
				london & 0, 0, 0, 1 & 0.53 & 18, 12, -11 \\
				berlin & 0.0299, 0, -0.0599, 0.9978 & 0.8 & -1, 1.8, -1.3 \\
				alameda & -0.1867, 0, 0, 0.9824 & 0.64 & 3, 2.6, -1 \\
				\bottomrule
			\end{tabular}%
		}
		\vspace{-1em}
	\end{table}
	
	\vspace{0.3em}
	\noindent\textbf{Scene Scale Calibration.}
	Another critical limitation of raw, trained 3DGS scenes is the absence of an intrinsic real-world scale--also due to COLMAP. 
	When rendered stereoscopically in a VR headset, this lack of calibration between scene units and physical world units can severely distort perceived object size.
	For example, under-scaled scenes can cause the users to feel disproportionately large, like a giant.
	
	This occurs because the physical inter-pupillary distance (IPD) becomes effectively magnified relative to the virtual world's scale. 
	Since VR rendering inherently relies on accurate simulation of binocular disparity between the user's eyes, calibrating the scene scale is essential for preserving immersion and visual comfort.
	
	Similar to scene tilt correction, we avoid directly modifying the trained 3DGS \texttt{.ply} files. 
	Instead, we apply a per-scene \texttt{scale} factor at runtime that maps movements in real-world metric measurements to the scene's virtual units.
	Using the Blende-KIRI add-on workflow, we introduced dimensionally accurate reference objects, e.g., a 1-meter cube, into each scene. 
	By comparing known dimensions from the real scene (e.g., width of a vehicle, rise height of staircases) to their 3DGS representations, we iteratively adjusted the scene scale within Blender's unit system. 
	These calibrated scale factors were recorded and applied during runtime, ensuring the perceived virtual scene conforms to real-world proportions and supports perceptually correct IPD rendering for stereo vision.

	\vspace{0.3em}
	\noindent\textbf{Initial View Positioning.}
	Besides tilt and scale, the initial view position also influences the user's first impression and subsequent exploration. 
	The default origin ($[0,0,0]$) of a 3DGS scene often corresponds to the center of the captured volume, which can result in undesirable starting viewpoints, such as inside a tree or a wall or floating in the air. 
	To improve user experiences, we manually selected semantically meaningful and physically plausible initial camera positions for each scene--typically floor-level regions with ample surrounding navigability. These locations were chosen to emulate natural human perspective, facilitate intuitive exploration, and avoid immediate occlusions or collisions.
	
	\subsection{Record-n-Replay Features}
	
	We extend the SIBR core rendering engine with record-and-replay features tailored for 3DGS in OpenXR.
	
	\vspace{0.3em}
	\noindent\textbf{The Record Mode.}
	In the record mode, our modified OpenXR module captures fine-grained, per-frame data during a user's VR session. 
	For each rendered frame and for each eye (left and right), we record a comprehensive set of parameters:
	the field-of-view (FOV), eye position, head orientation (as a quaternion), as well as the eye gaze position and gaze orientation quaternion.
	The data is synchronized with the rendering loop and saved to a structured \texttt{csv} file with precise timing.
	This enables a complete and accurate offline reconstruction of user's viewpoints.
	
	\vspace{0.3em}
	\noindent\textbf{The Replay Mode.}
	The replay mode leverages the recorded traces to reproduce the original VR session for analysis and rendered view generation. 
	During replay, the recorded trace is parsed and the logged data is injected line-by-line into the rendering pipeline, overriding the HMD pose information. 
	Internally, this substitution is handled via the \texttt{loadViewData()} method, which deserializes the \texttt{csv} traces and updates the per-frame \texttt{ViewData} structure prior to rendering. 
	
	To generate video output, rendered frames are captured directly from GPU memory using OpenGL's \texttt{glGetTexImage()} API.
	These frames are then converted into an OpenCV-compatible format for efficient encoding. 
	The replay mode generates two separate videos, one for each eye, that precisely replicate the original stereoscopic experience.
	
	\subsection{Data Output Format}
	
	Table \ref{tbl:format} outlines the structure of the recorded \texttt{csv} traces.
	To support stereoscopic views, each rendered frame contains the rendered views for both left and right eyes, distinguished by ``ViewIndex'' column (0 for left, and 1 for right).
	The FOV of each eye is captured by \texttt{FOV1}, \texttt{FOV2}, \texttt{FOV3}, and \texttt{FOV4}, representing its left, right, top, and bottom in radians, respectively. 
	Due to the IPD, the left and right eyes see views in different world coordinates.
	Thus, \texttt{Pos$\_\texttt{X}$, Pos$\_\texttt{Y}$, Pos$\_\texttt{Z}$}, representing view/head positions would be different for the left and right eyes. 
	On the other hand, the quaternions of the tracked head orientation in the world space are given in 
	\texttt{Quat$\_\texttt{X}$}, \texttt{Quat$\_\texttt{Y}$}, \texttt{Quat$\_\texttt{Z}$}, and \texttt{Quat$\_\texttt{W}$}, which would be the same for the left and right eyes.  
	
	The recorded traces also contain user gaze information
	if supported by the headset (e.g., Meta Quest Pro).
	Here, \texttt{GazePos$\_\texttt{X}$, GazePos$\_\texttt{Y}$, GazePos$\_\texttt{Z}$} represent the \textbf{eye gaze position}, in world space.
	These values will be very similar to, but distinct from, the corresponding \textbf{eye position} \texttt{Pos$\_\texttt{X}$, Pos$\_\texttt{Y}$, Pos$\_\texttt{Z}$}. 
	\texttt{GazeQ$\_\texttt{X}$}, \texttt{GazeQ$\_\texttt{Y}$}, \texttt{GazeQ$\_\texttt{Z}$}, and \texttt{GazeQ$\_\texttt{W}$} provide the \textbf{eye gaze orientation} in quaternion, in the world space. These columns can differ significantly from the head orientation due to eye movement within the eye sockets. 
	Finally, we record relative timestamps in milliseconds for each recorded frame.

	Table \ref{tbl:sample_rounded} gives sample columns from two rendered frames. We note that every two consecutive rows correspond to the views for the left and right eye view of each rendered frame.

	\begin{table}[!t] 
		\caption{Columns of the Recorded User Traces in \texttt{csv}}
		\vspace{-0.5em}
		\label{tbl:format}
		\resizebox{0.45\textwidth}{!}{%
			\begin{tabularx}{0.45\textwidth}{@{} l >{\raggedright\arraybackslash}X @{}}
				\toprule
				{\bf Name} & {\bf Description} \\
				\midrule
				ViewIndex & Left eye: 0; right eye: 1. \\
				FOV1 (rad) & The left FOV angle. \\
				FOV2 (rad) & The right FOV angle. \\
				FOV3 (rad) & The top FOV angle. \\
				FOV4 (rad) & The bottom FOV angle. \\
				Position X,Y,Z & Eye position (i.e., head position offset by half of IPD) in the world space. \\
				Quaternion X,Y,Z,W & Head/view orientation as a quaternion, in the world space (same for left and right eyes). \\
				GazePos X,Y,Z & Eye gaze position in the world space. \\
				GazeQ X,Y,Z,W & Eye gaze orientation as a quaternion, in the world space. \\
				Timestamp & Time offset in \texttt{ms} since the left eye of the first frame is recorded.\\
				\bottomrule
			\end{tabularx}
		}
	\end{table}
	
	\begin{table*}[!t]
		\centering
		\caption{Sample Values Extracted from a Recorded \texttt{csv}}
		\vspace{-0.5em}
		\label{tbl:sample_rounded}
		\resizebox{\textwidth}{!}{%
			\begin{tabular}{
					S[table-format=1.0]
					S[table-format=-1.3, round-mode=places, round-precision=3]   
					S[table-format=1.3, round-mode=places, round-precision=3]    
					S[table-format=-1.3, round-mode=places, round-precision=3]   
					S[table-format=1.3, round-mode=places, round-precision=3]    
					S[table-format=-1.3, round-mode=places, round-precision=3]   
					S[table-format=-1.3, round-mode=places, round-precision=3]   
					S[table-format=1.3, round-mode=places, round-precision=3]    
					S[table-format=1.3, round-mode=places, round-precision=3]    
					S[table-format=1.3, round-mode=places, round-precision=3]    
					S[table-format=1.3, round-mode=places, round-precision=3]    
					S[table-format=1.3, round-mode=places, round-precision=3]    
					S[table-format=1.3, round-mode=places, round-precision=3]    
					S[table-format=1.3, round-mode=places, round-precision=3]    
					S[table-format=1.3, round-mode=places, round-precision=3]    
					S[table-format=1.3, round-mode=places, round-precision=3]    
					S[table-format=-1.3, round-mode=places, round-precision=3]   
					S[table-format=-1.3, round-mode=places, round-precision=3]   
					S[table-format=1.3, round-mode=places, round-precision=3]    
				}
				\toprule
				{\bf View} & {\bf FOV1} & {\bf FOV2} & {\bf FOV3} & {\bf FOV4} & {\bf Pos$_X$} & {\bf Pos$_Y$} & {\bf Pos$_Z$} & {\bf Quat$_X$} & {\bf Quat$_Y$} & {\bf Quat$_Z$} & {\bf Quat$_W$} & {\bf GazeQ$_X$} & {\bf GazeQ$_Y$} & {\bf GazeQ$_Z$} & {\bf GazeQ$_W$} & {\bf GazePos$_X$} & {\bf GazePos$_Y$} & {\bf GazePos$_Z$} \\
				{\bf Index} & {\bf (rad)} & {\bf (rad)} & {\bf (rad)} & {\bf (rad)} & {} & {} & {} & {} & {} & {} & {} & {} & {} & {} & {} & {} & {} & {} \\ 
				\midrule
				0 & -0.942478 & 0.698132 & -0.942478 & 0.733038 & -3.66908 & -3.65709 & 4.65788 & 0.494687 & 0.294258 & 0.123821 & 0.808310 & 0.250753 & 0.0845578 & 0.0237413 & 0.964059 & -3.66845 & -3.65671 & 4.65700 \\
				1 & -0.698132 & 0.942478 & -0.942478 & 0.733038 & -3.51258 & -3.56052 & 4.58845 & 0.494687 & 0.294258 & 0.123821 & 0.808310 & 0.245048 & 0.1037840 & 0.0453922 & 0.962871 & -3.51320 & -3.56090 & 4.58873 \\
				0 & -0.942478 & 0.698132 & -0.942478 & 0.733038 & -3.66901 & -3.65635 & 4.65733 & 0.494082 & 0.293893 & 0.122980 & 0.808941 & 0.248543 & 0.0871354 & 0.0263629 & 0.964333 & -3.66845 & -3.65617 & 4.65724 \\
				1 & -0.698132 & 0.942478 & -0.942478 & 0.733038 & -3.51234 & -3.56015 & 4.58775 & 0.494082 & 0.293893 & 0.122980 & 0.808941 & 0.242903 & 0.1064400 & 0.0481759 & 0.962989 & -3.51307 & -3.56065 & 4.58826 \\
				\bottomrule
			\end{tabular}%
		}
	\end{table*}

	\section{\Lname\ Dataset}
	\label{sec:dataset}
	\subsection{Data Collection Methodology}
	
	\noindent\textbf{Sites.}
	We collected user navigation traces at two sites: Rutgers University (RU) in the New Jersey, USA and National Tsing Hua University (NTHU) in Hsin-Chu, Taiwan. 
	The data collection protocols at both institutions received approval from their respective Institutional Review Boards (IRB).
	
	\noindent\textbf{Participants.}
	A total of 46 participants were recruited across the two sites: 22 at RU and 24 at NTHU. The participants' ages ranged from 18 to 70. 
	
\begin{figure}[!t]
		\centering 
		\begin{subfigure}[b]{0.225\textwidth}
			\centering
			\includegraphics[width=\textwidth]{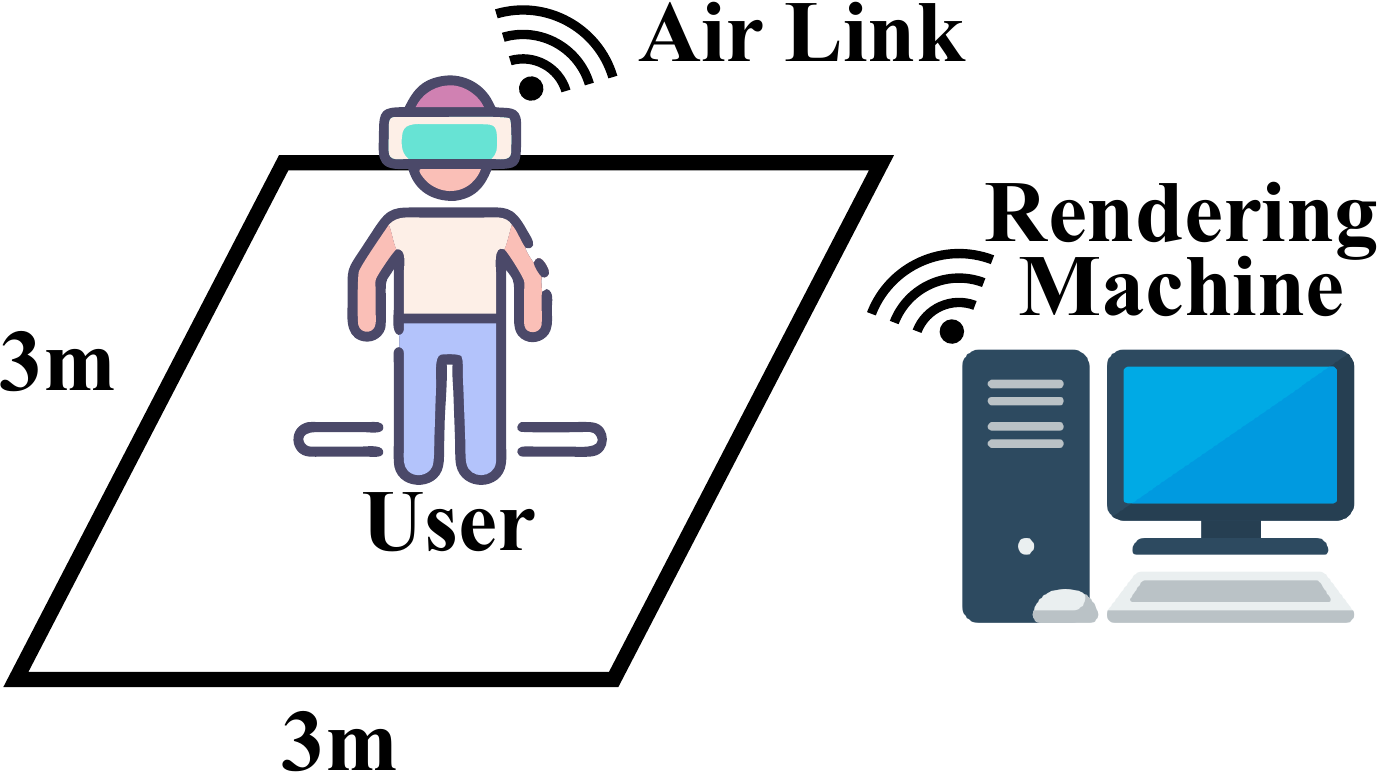}
			\caption{Untethered user trace collection setup at the RU site.}
			\label{fig:sub1}
		\end{subfigure}
		\hfill 
		\begin{subfigure}[b]{0.225\textwidth}
			\centering
			\includegraphics[width=\textwidth]{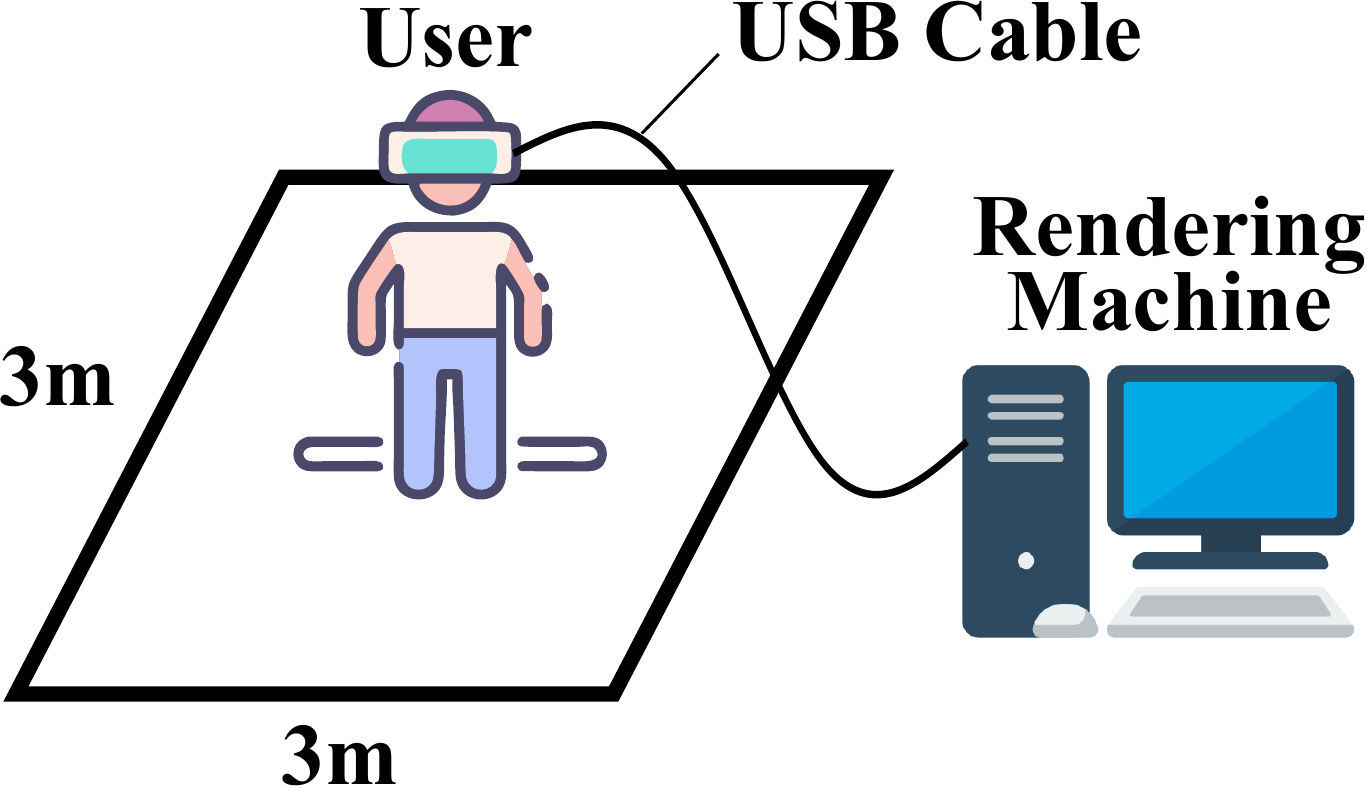}
			\caption{Tethered user trace collection setup at the NTHU site.}
			\label{fig:sub2}
		\end{subfigure}
		\vspace{-1em}
		\caption{Real-time user trace collection setup showing the tracked space, user with HMD, HMD connection, and the rendering machine.}
		\vspace{-1em}
		\label{fig:env_setup}
	\end{figure}

	\noindent\textbf{Apparatus.}
	The experiment setup specifications at both collection sites are listed in Table \ref{tbl:setups}.
	At both locations, participants used a Meta Quest Pro headset with eye gaze tracking enabled and navigated a \threemeterarea physical play area corresponding to the OpenXR ``Stage Space.'' 
	The main differences between the two sites were the HMD connection methods and rendering GPUs. 
	Specifically, as shown in Figure \ref{fig:env_setup}, the RU site provided an untethered VR experience using Meta Air Link, powered by an Nvidia RTX 4090 GPU; while the NTHU site used a tethered USB Link cable connected to a desktop PC with an Nvidia RTX 3080 Ti GPU.

	\begin{table}[!t]
		\centering
		\caption{Dataset Collection Setups at the RU and NTHU Sites}
		\vspace{-0.5em}
		\label{tbl:setups}
		\resizebox{0.48\textwidth}{!}{
			\begin{tabular}{lll}
				\toprule
				\textbf{Component} & \textbf{RU Setup} & \textbf{NTHU Setup} \\
				\midrule
				HMD & Meta Quest Pro & Meta Quest Pro\\\addlinespace
				Rendering & GPU: Nvidia RTX 4090 & GPU: NVIDIA RTX 3080 Ti \\
				Machine   & CPU: Intel i9-14900KF & CPU: Intel i9-9920X \\
				\addlinespace 
				Connection        & Wireless (Meta Air Link) & Wired (5-meter USB Link) \\
				\addlinespace
				Area & \threemeterarea & \threemeterarea \\\addlinespace
				OS & Windows 11 & Windows 10 \\\addlinespace
				\midrule
				Participants & 22 & 24 \\
				\bottomrule
			\end{tabular}
		}
	\end{table}

\begin{figure}[!t]
		\centering
		\begin{subfigure}[b]{0.11\textwidth}
			\includegraphics[width=\textwidth]{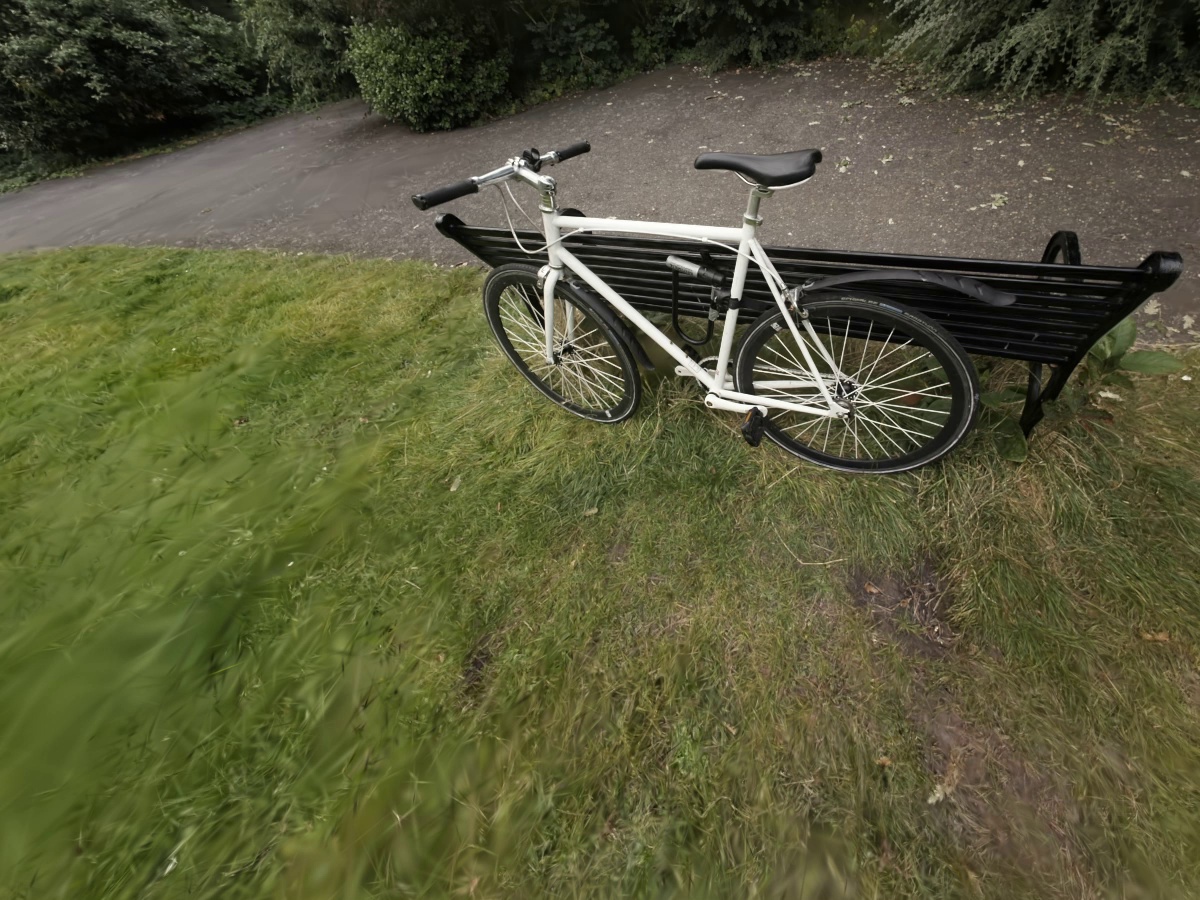}
			\caption{bicycle}
		\end{subfigure}
		\begin{subfigure}[b]{0.11\textwidth}
			\includegraphics[width=\textwidth]{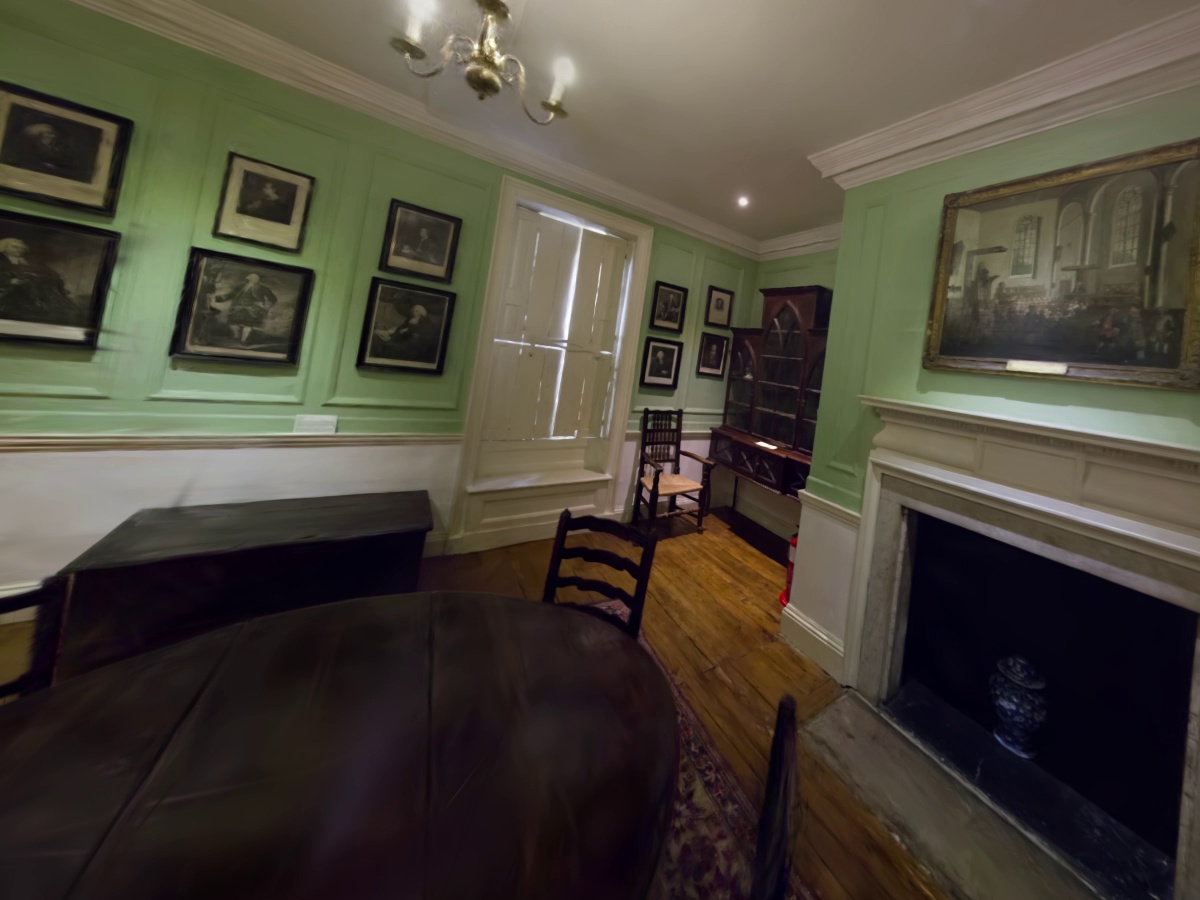}
			\caption{drjohnson}
		\end{subfigure}
		\begin{subfigure}[b]{0.11\textwidth}
			\includegraphics[width=\textwidth]{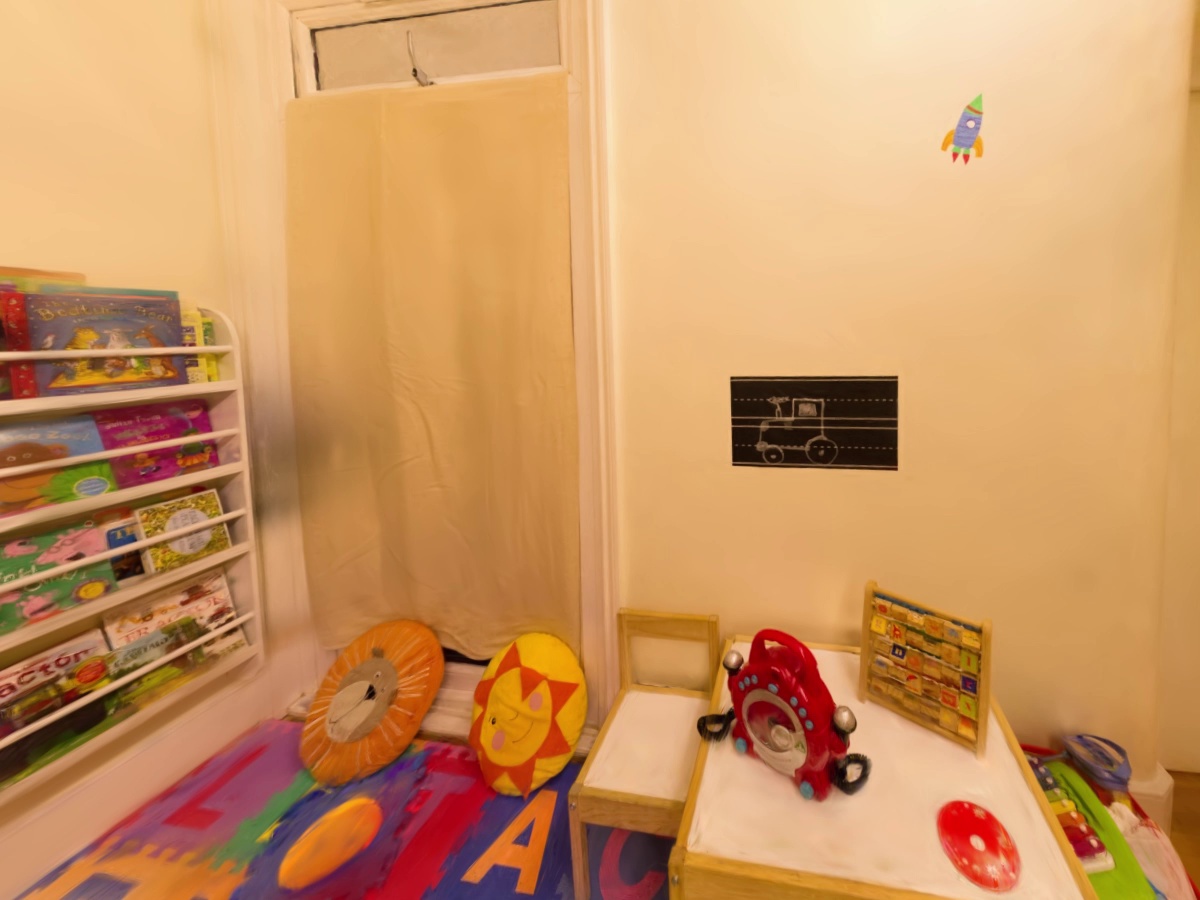}
			\caption{playroom}
		\end{subfigure}
		\begin{subfigure}[b]{0.11\textwidth}
			\includegraphics[width=\textwidth]{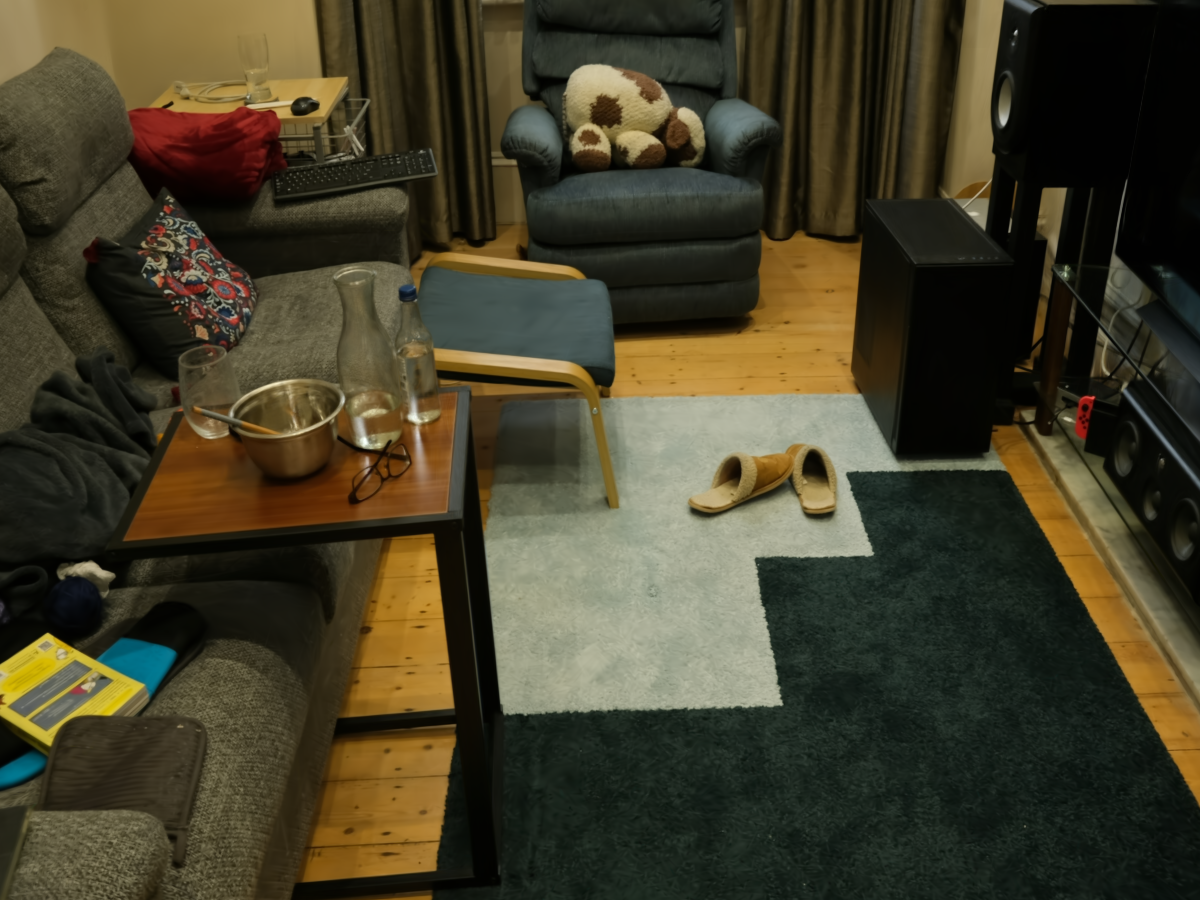}
			\caption{room}
		\end{subfigure}
		\begin{subfigure}[b]{0.11\textwidth}
			\includegraphics[width=\textwidth]{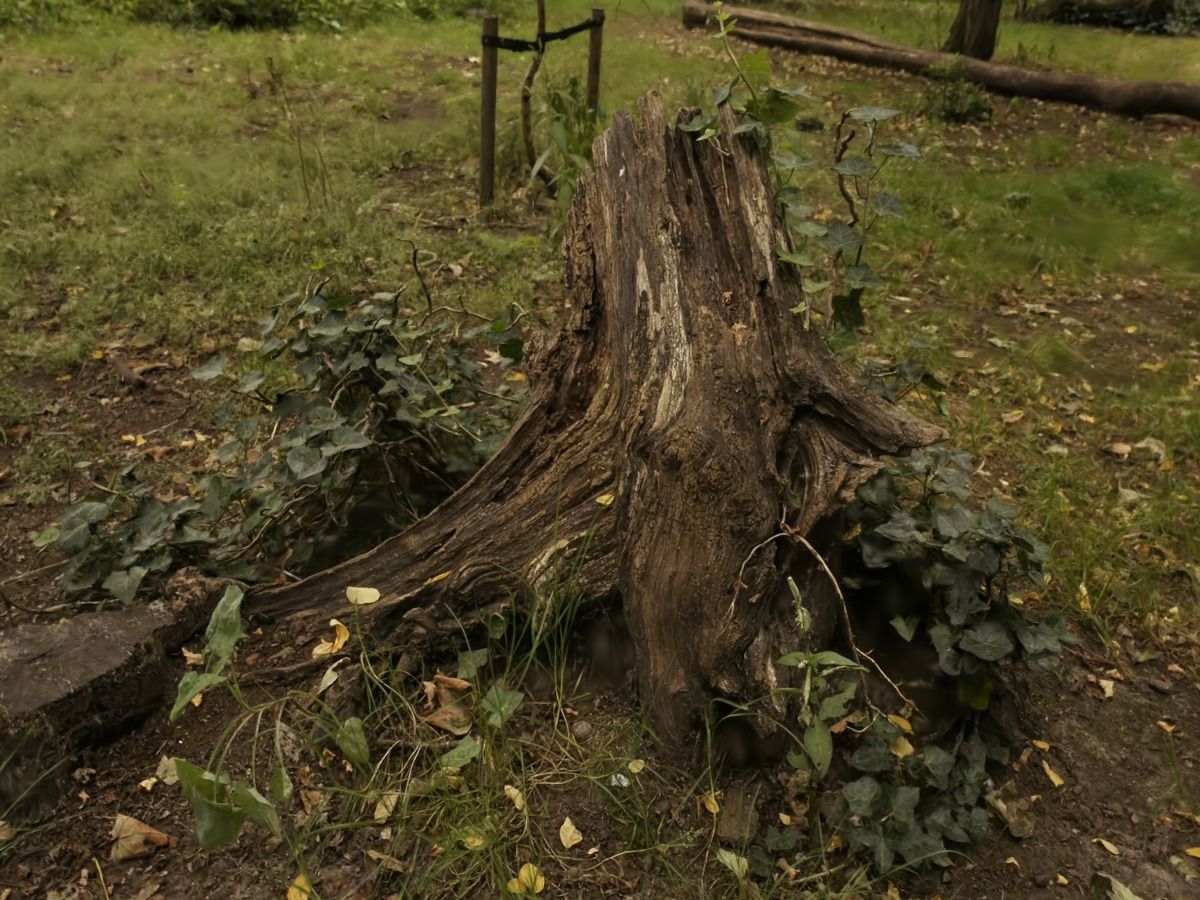}
			\caption{stump}
		\end{subfigure}
		\begin{subfigure}[b]{0.11\textwidth}
			\includegraphics[width=\textwidth]{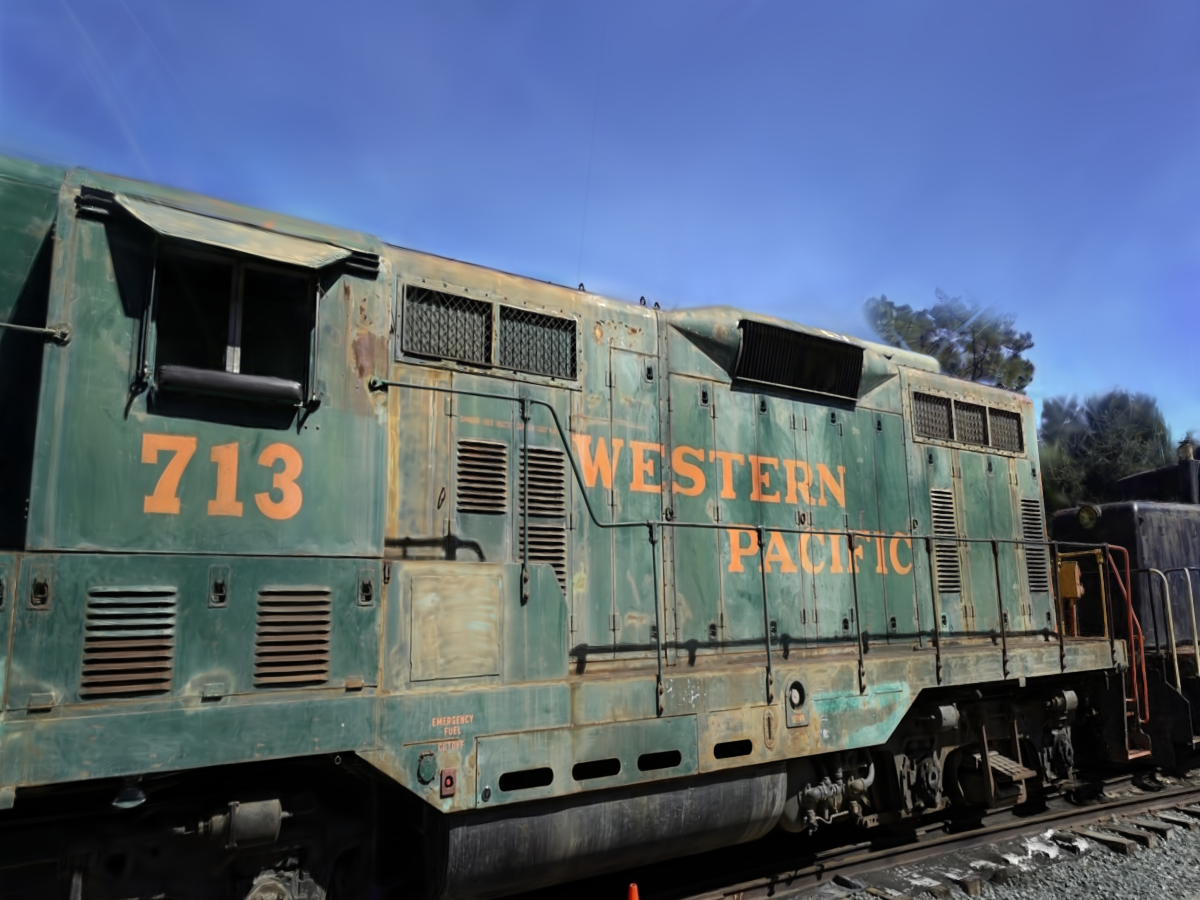}
			\caption{train}
		\end{subfigure}
		\begin{subfigure}[b]{0.11\textwidth}
			\includegraphics[width=\textwidth]{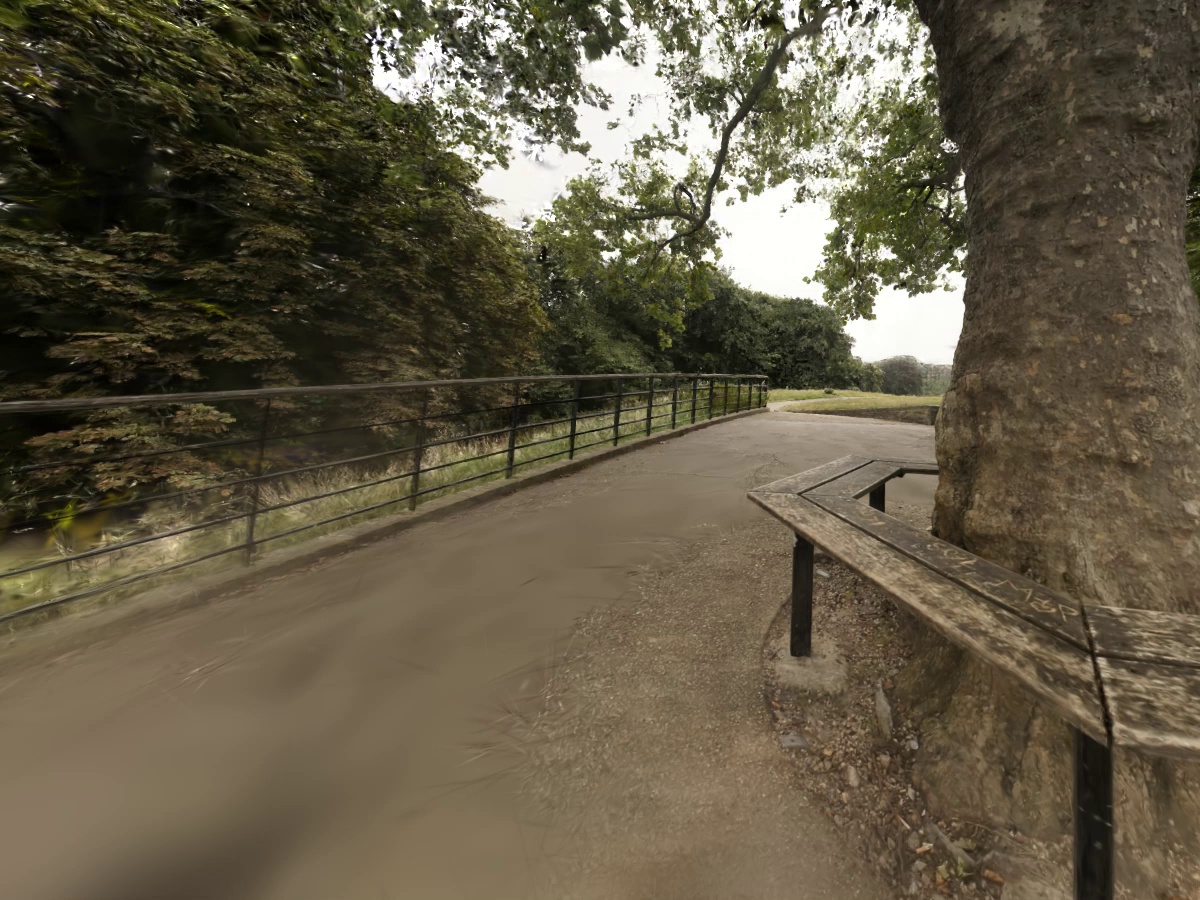}
			\caption{treehill}
		\end{subfigure}
		\begin{subfigure}[b]{0.11\textwidth}
			\includegraphics[width=\textwidth]{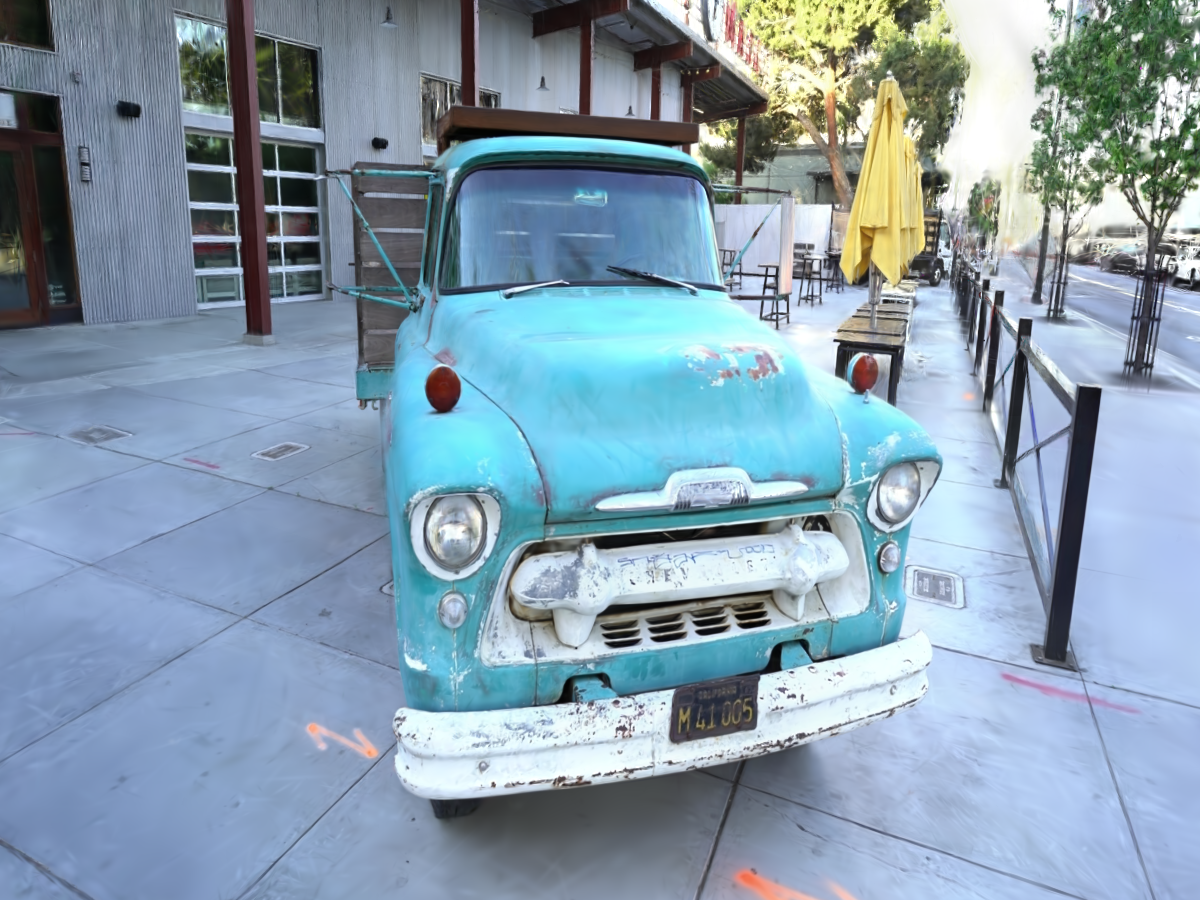}
			\caption{truck}
		\end{subfigure}
		\vspace{-1em}
		\caption{Pre-trained 3DGS scenes from the 3DGS paper~\cite{kerbl20233d}, used for user navigation in our dataset.}
		\label{fig:pretrained_3dgs}
	\end{figure}

	\noindent\textbf{Stimuli.}
	Stereoscopic views are rendered at $2160\times 2224$ resolution per eye. 
	Twelve distinct real-world scenes were used for user navigation. 
	Eight of these scenes were selected from the twelve pre-trained 3DGS scenes presented in the original 3DGS paper~\cite{kerbl20233d}, shown in Figure \ref{fig:pretrained_3dgs}. These scenes originate from three datasets: the MipNeRF360 dataset~\cite{barron2022mip}, the Tanks\&Temples dataset~\cite{knapitsch2017tanks}, and the Deep Blending dataset~\cite{hedman2018deep}. 
	They contain a variety of real-world scenes including indoor, outdoor, and natural environments. 
	For our study, we used the ``garden'' scene from this pre-trained dataset for participant training. We excluded ``kitchen'', ``flowers'', ``counter'', and ``bonsai'' scenes due to their limited viewing volumes, making them unsuitable for VR navigation.
	
\begin{figure}[!t]
		\centering
		\begin{subfigure}[b]{0.11\textwidth}
			\includegraphics[width=\textwidth]{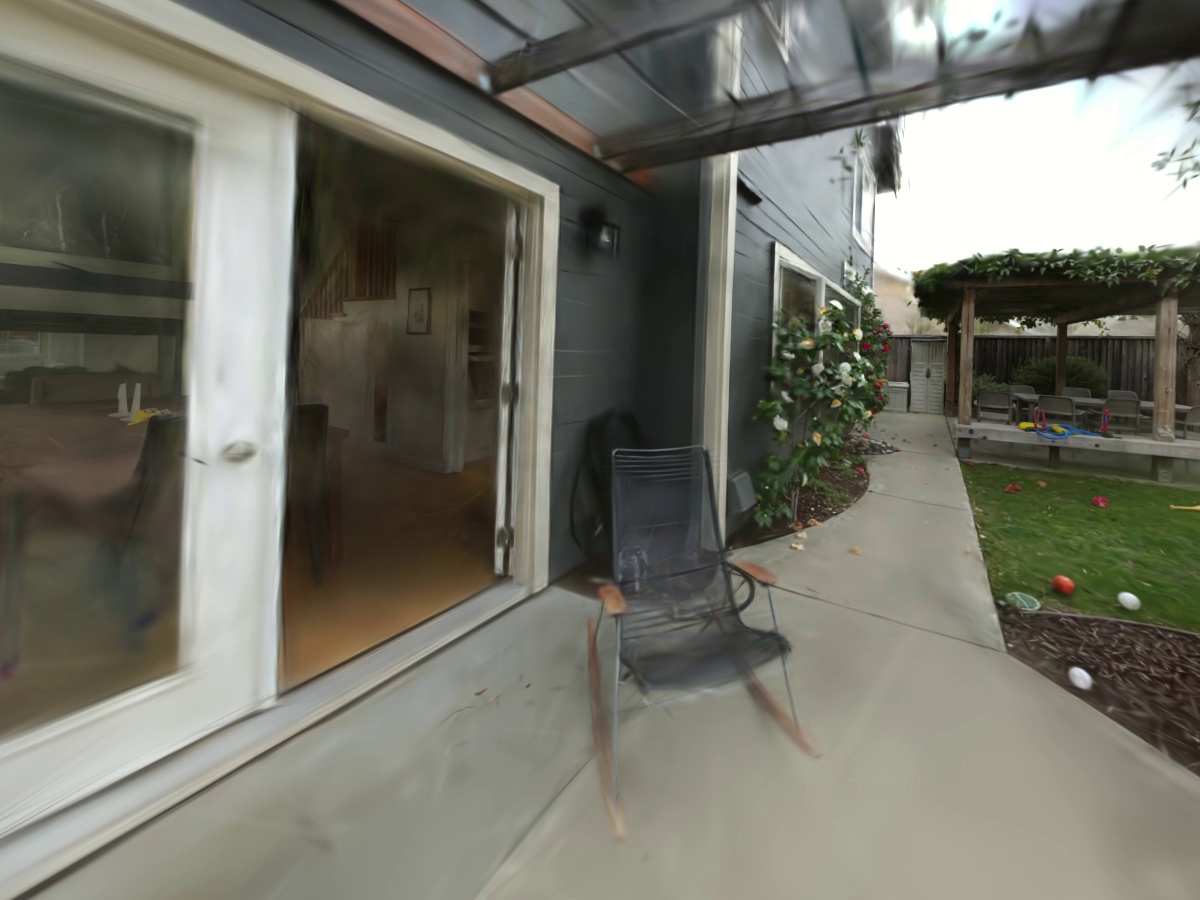}
			\caption{alameda}
		\end{subfigure}
		\begin{subfigure}[b]{0.11\textwidth}
			\includegraphics[width=\textwidth]{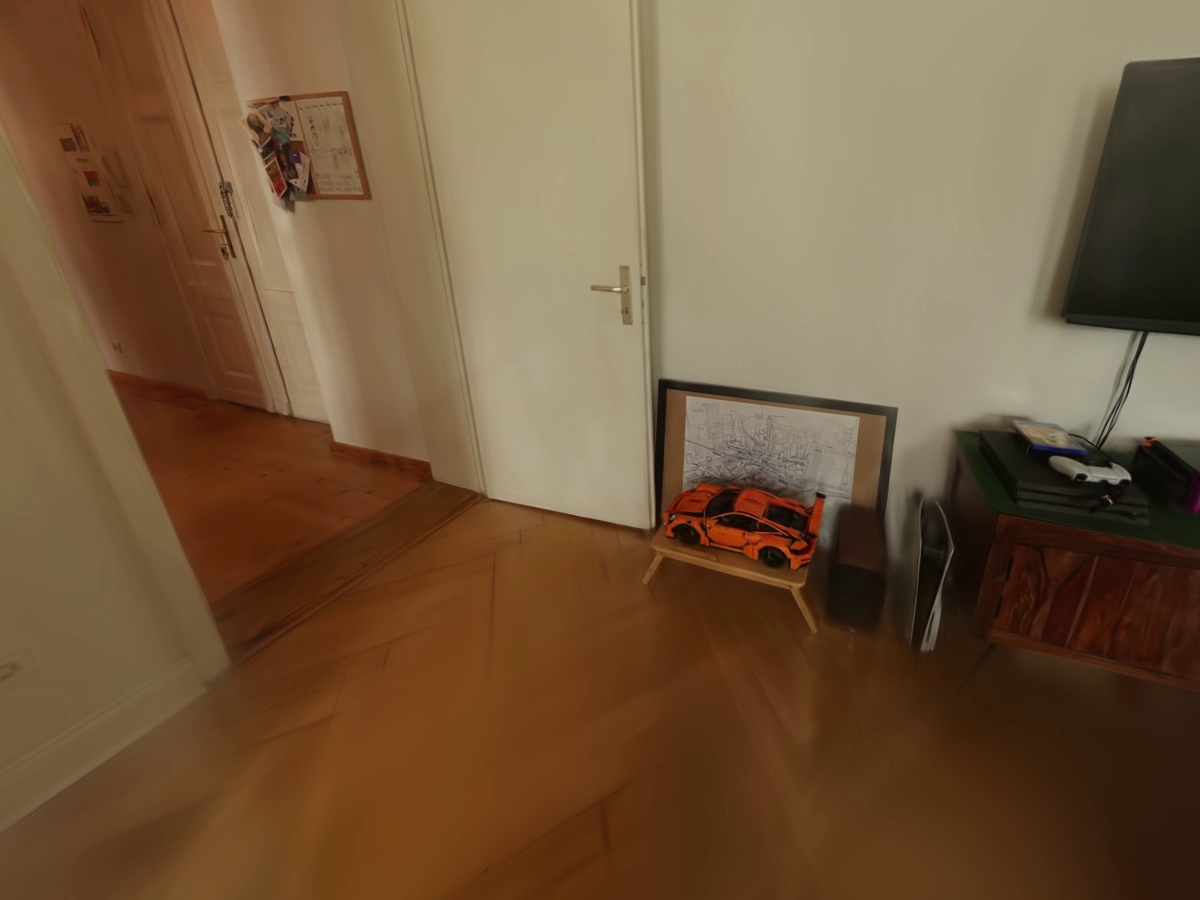}
			\caption{berlin}
		\end{subfigure}
		\begin{subfigure}[b]{0.11\textwidth}
			\includegraphics[width=\textwidth]{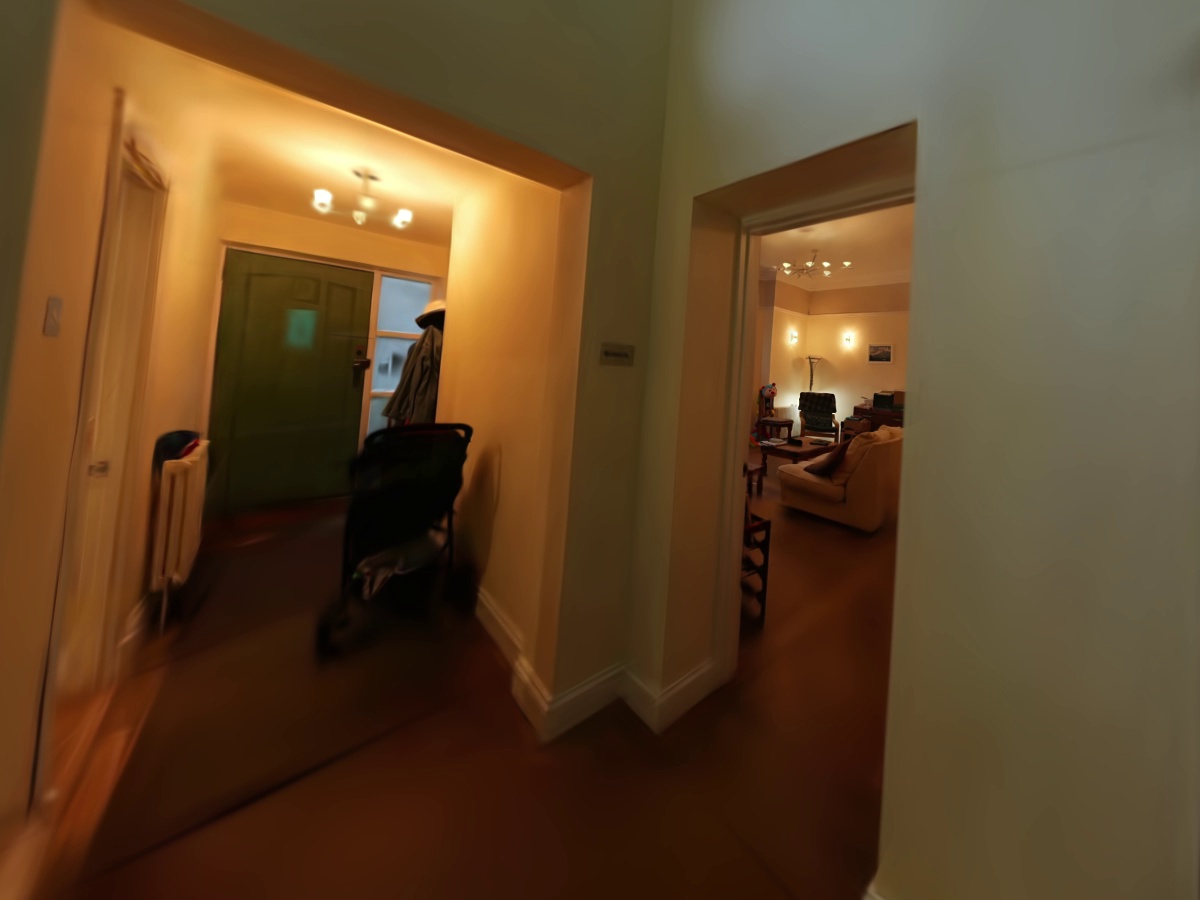}
			\caption{london}
		\end{subfigure}
		\begin{subfigure}[b]{0.11\textwidth}
			\includegraphics[width=\textwidth]{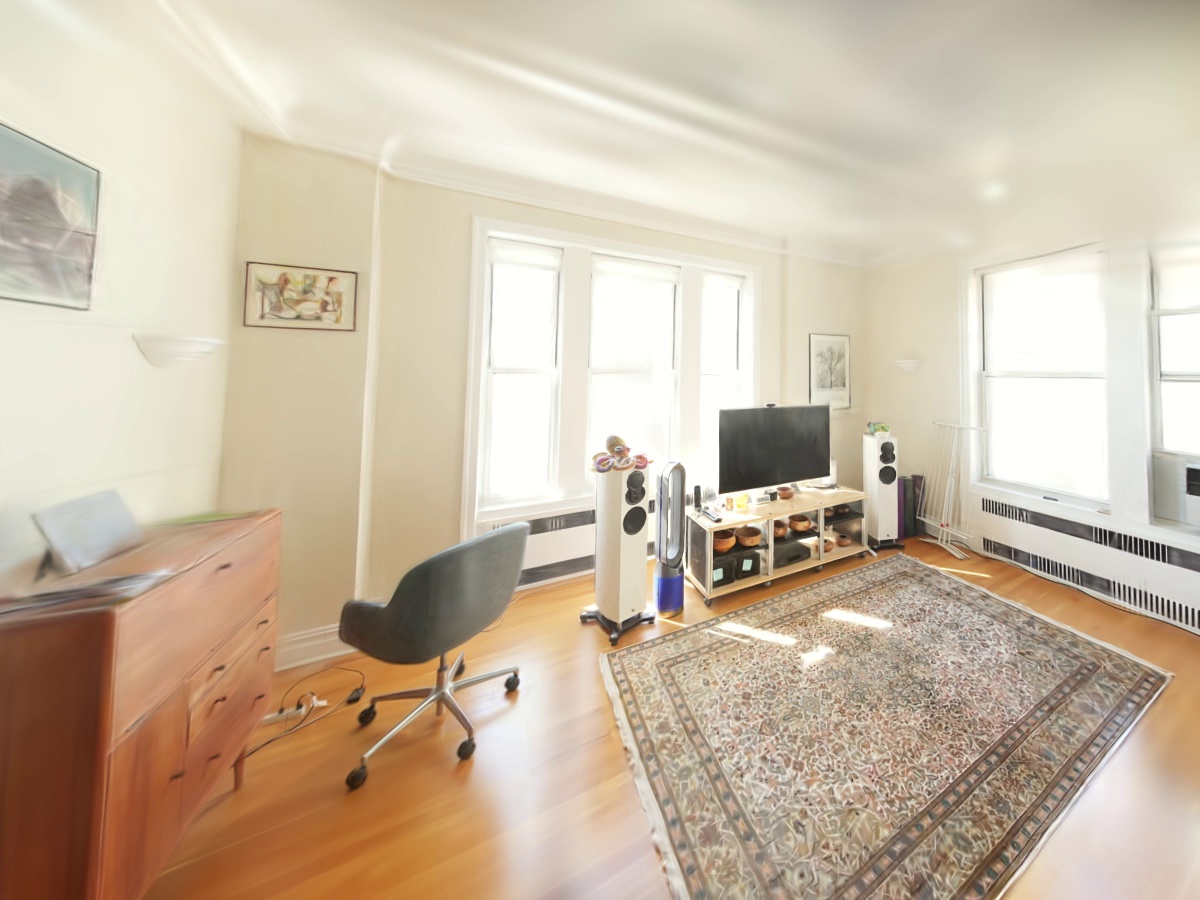}
			\caption{nyc}
		\end{subfigure}
		\vspace{-1em}
		\caption{Newly trained 3DGS scenes from the ZipNeRF dataset~\cite{barron2023zip,duckworth2024smerf}, included to expand the diversity of our scene collection.}
		\vspace{-1em}
		\label{fig:zipnerf_scenes}
	\end{figure}

	In addition, we also trained 3DGS representations for four scenes from the ZipNeRF dataset~\cite{barron2023zip,duckworth2024smerf}, namely ``alameda'', ``berlin'', ``london'', and ``nyc'', shown in Figure \ref{fig:zipnerf_scenes}. These scenes are mainly indoor scenes, while some of them also feature outdoor sections.

	\noindent\textbf{Procedure.}
	Each participants explored twelve scenes for one minute each, in the \texttt{free world standing} (\texttt{fws}) mode. 
	Participants were instructed to freely explore the virtual scene via natural movements such as walking and turning in the \threemeterarea area. No specific task was assigned to the participants during the sessions.
	A one-minute break was provided between each scene exploration to allow for participant rest and for the system to load the next scene.
	
	\subsection{Folder Structure}
	Our dataset is organized in a hierarchical folder structure:
	
	\dirtree{%
		.1 dataset/.
		.2 truck/.
		.3 user101\_truck.csv.
		.3 user102\_truck.csv.
		.3 ....
		.2 alameda/.
		.3 user101\_alameda.csv.
		.3 ....
	}

	Each top-level folder corresponds to the specific 3DGS scenes (e.g., ``train'', ``truck''). 
	Within each scene folder, each user trace is stored as a \texttt{csv} file named as:
	\texttt{\{user\}\_\{scene\}.csv}.
	For instance, \texttt{user1\_truck.csv} records the trace of user1 exploring the ``truck'' scene.

	\section{\Lname\ Utility Tools}
	\label{sec:utility}
	We developed a suite of utilities for the  interoperability, reproducibility, and visualization of the collected traces. We include these utilities in the \name\ software's \texttt{utils} folder and briefly describe them in this section. 
	
	\vspace{0.3em}
	\noindent\textbf{Conversion from Virtual World Coordinates to Physical Stage Coordinates.}
	The \name\ dataset records user navigation traces including the head pose and eye gaze information in ``virtual world coordinates''. This facilitates direct replay of the traces for view generation. 
	We provide a utility to convert these traces to ``physical stage coordinates'', which represent the user's movements in the physical world on a 1:1 metric scale (i.e., a one-meter physical movement in corresponds to a one-unit displacement in these coordinates). 
	This conversion undoes the scene initialization transforms including the scene tilt correction, scene scaling, and initial view positioning, detailed in Section \ref{sec:initialization}. 
	The resulting ``physical stage coordinates'' effectively reconstruct user's movement in the \threemeterarea physical space. 
	We use the converted physical stage coordinates in our dataset analysis in Section \ref{sec:analyis}.
	
	\vspace{0.3em}
	\noindent\textbf{Compatibility with Other Frameworks.}
	The \name\ dataset stores user navigation traces in \texttt{.csv} format, recording camera positions and rotation quaternions. Popular modern viewers, such as the web viewer in NeRFstudio~\cite{nerfstudio} and SGSS~\cite{zhu2025sgss}, use \texttt{.json} format with a $4 \times 4$ homogeneous matrix for pose representation in a different coordinate system for trace replay. To enable compatibility with these frameworks, \name\ includes two utility tools: \texttt{csv2json.cpp} for converting recorded traces to \texttt{.json} format in the correct coordinate system and \texttt{json2csv.cpp} for convert \texttt{.json} files exported from other frameworks to \texttt{.csv} format, also in the correct coordinate system, enabling replay in the \name\ viewer. 
	With these two tools, \name\ ensures compatibility with other frameworks.

	\begin{figure}[!t]
		\centering
		\includegraphics[width=0.42\textwidth]{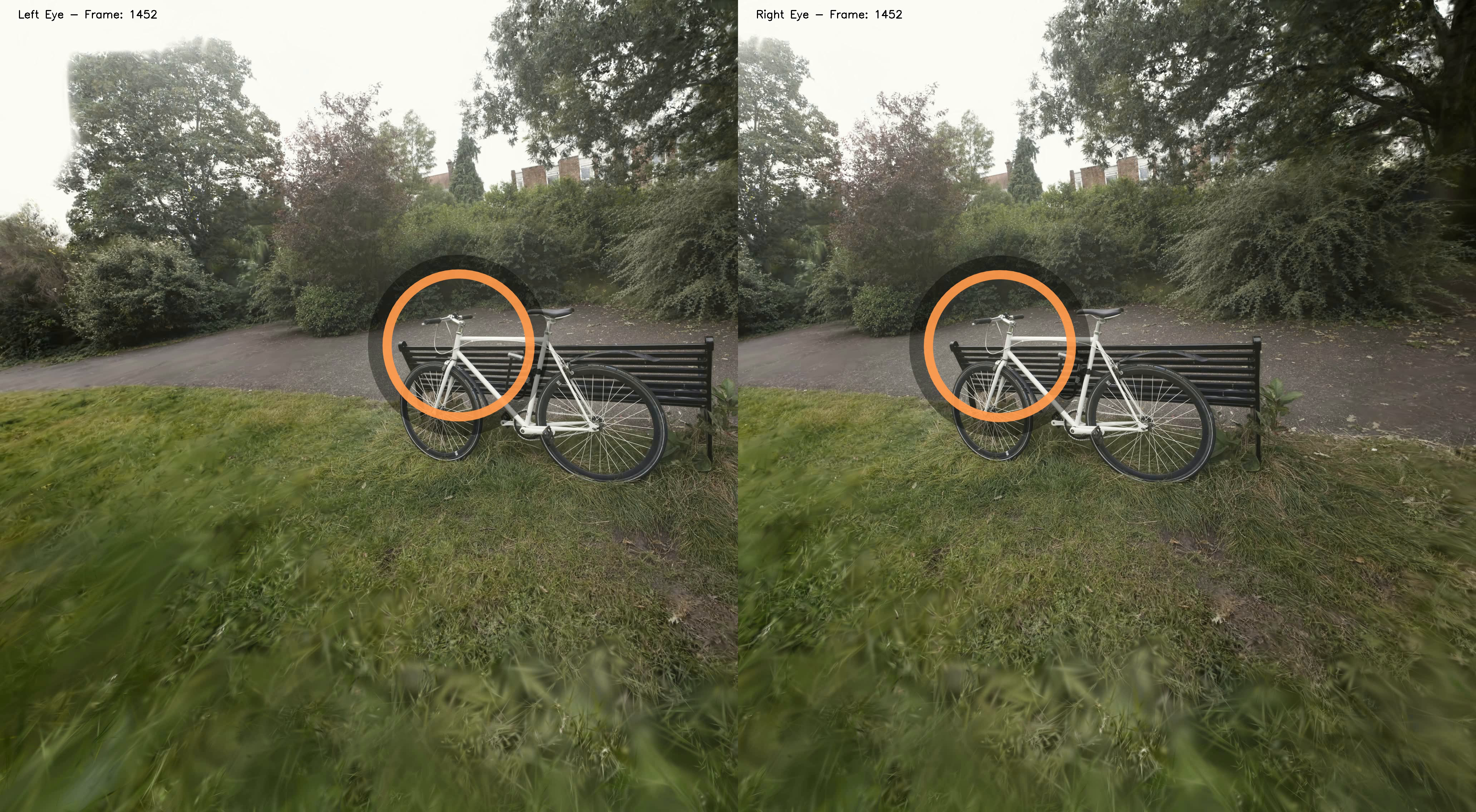}
		\vspace{-1em}
		\caption{Example eye gaze visualization of the bicycle scene.}
		\label{fig:eye-gaze}
	\end{figure}
	
	\vspace{0.3em}
	\noindent\textbf{Eye Gaze Visualization.}
	To visualize user attention during scene navigation, our eye gaze tool processes replayed stereoscopic videos by overlaying the recorded eye gaze data. 
	For each frame, a visual marker (e.g., a circle) is rendered on both left and right eye views to indicate the participant's fixation. The resulting video displays the stereoscopic view side-by-side, illustrating the user's gaze path. Figure \ref{fig:eye-gaze} shows an example frame with eye gaze overlaid on stereoscopic rendered views.

	\section{Dataset Analysis}
	\label{sec:analyis}
	
	\begin{figure}[!t] 
		\centering
		\begin{subfigure}[t]{0.225\textwidth}
			\centering
			\includegraphics[width=\linewidth]{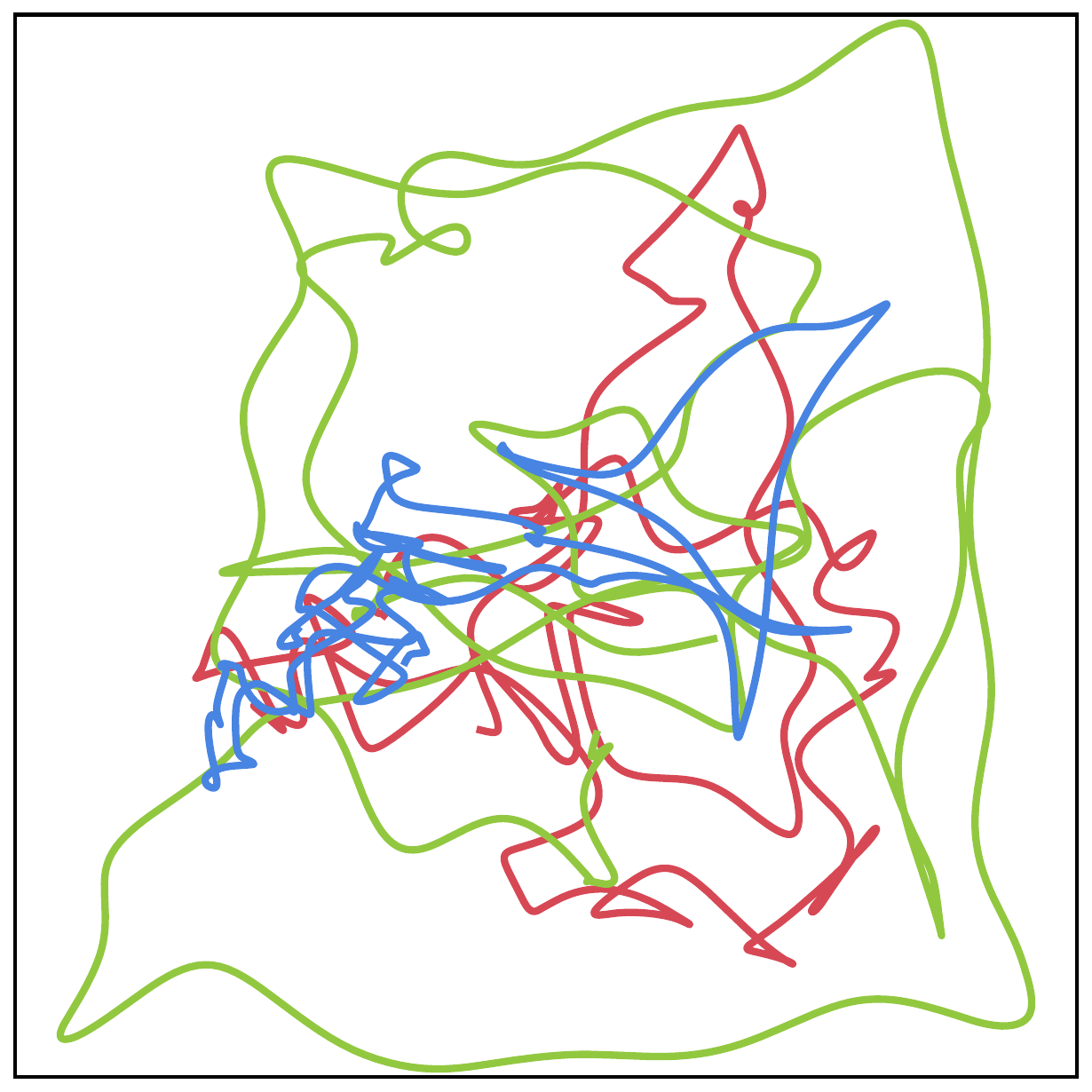} 
			\caption{bicycle $\mathrm{XZ}$-plane}
			\label{fig:scene1_xz}
		\end{subfigure}
		\hfill
		\begin{subfigure}[t]{0.225\textwidth}
			\centering
			\includegraphics[width=\linewidth]{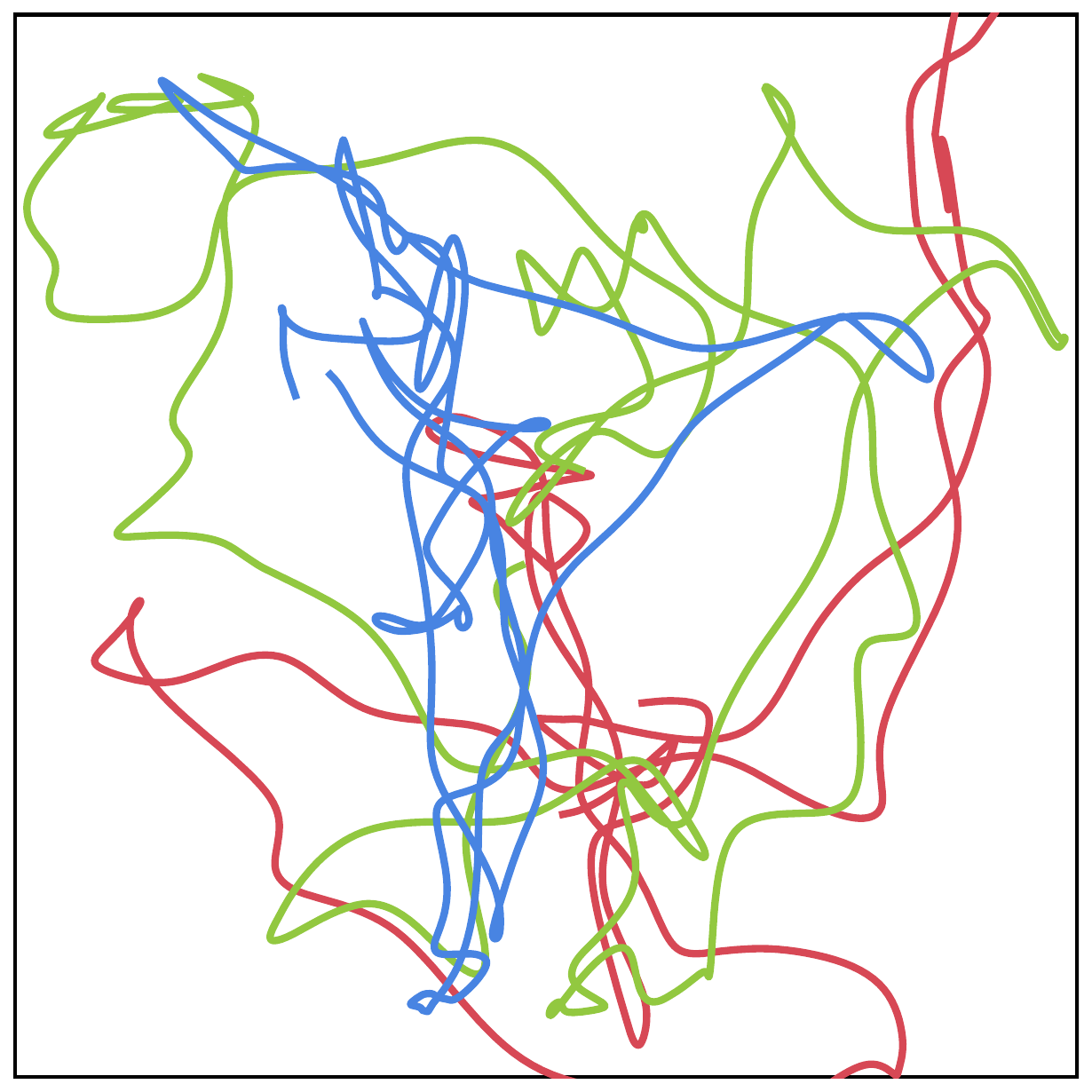} 
			\caption{nyc $\mathrm{XZ}$-plane}
			\label{fig:scene2_xz}
		\end{subfigure}
		
		\vspace{1em} 
		
		\begin{subfigure}[t]{0.225\textwidth}
			\centering
			\includegraphics[width=\linewidth]{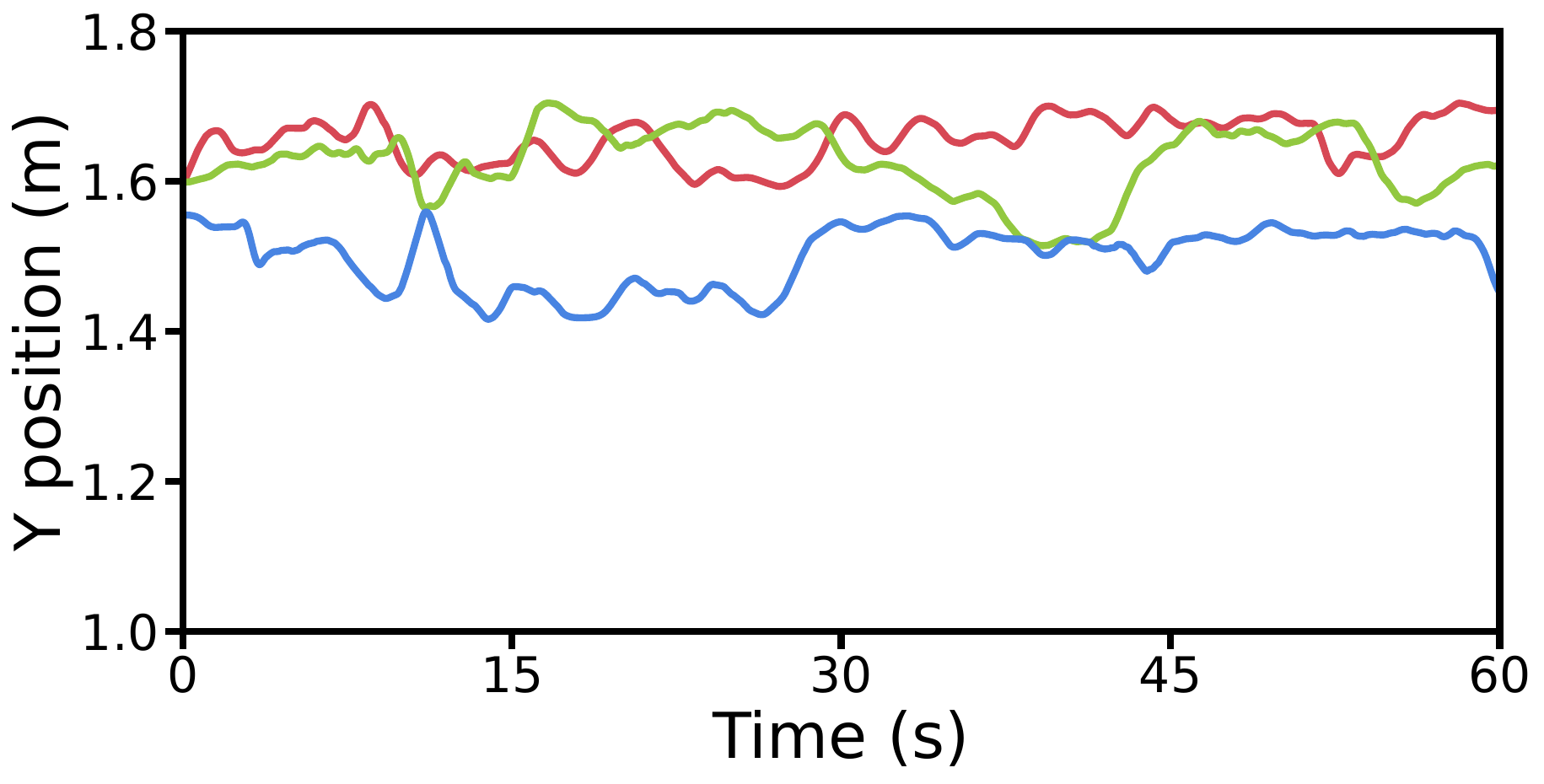} 
			\caption{bicycle $\mathrm{Y}$-axis}
			\label{fig:scene1_y}
		\end{subfigure}
		\hfill
		\begin{subfigure}[t]{0.225\textwidth}
			\centering
			\includegraphics[width=\linewidth]{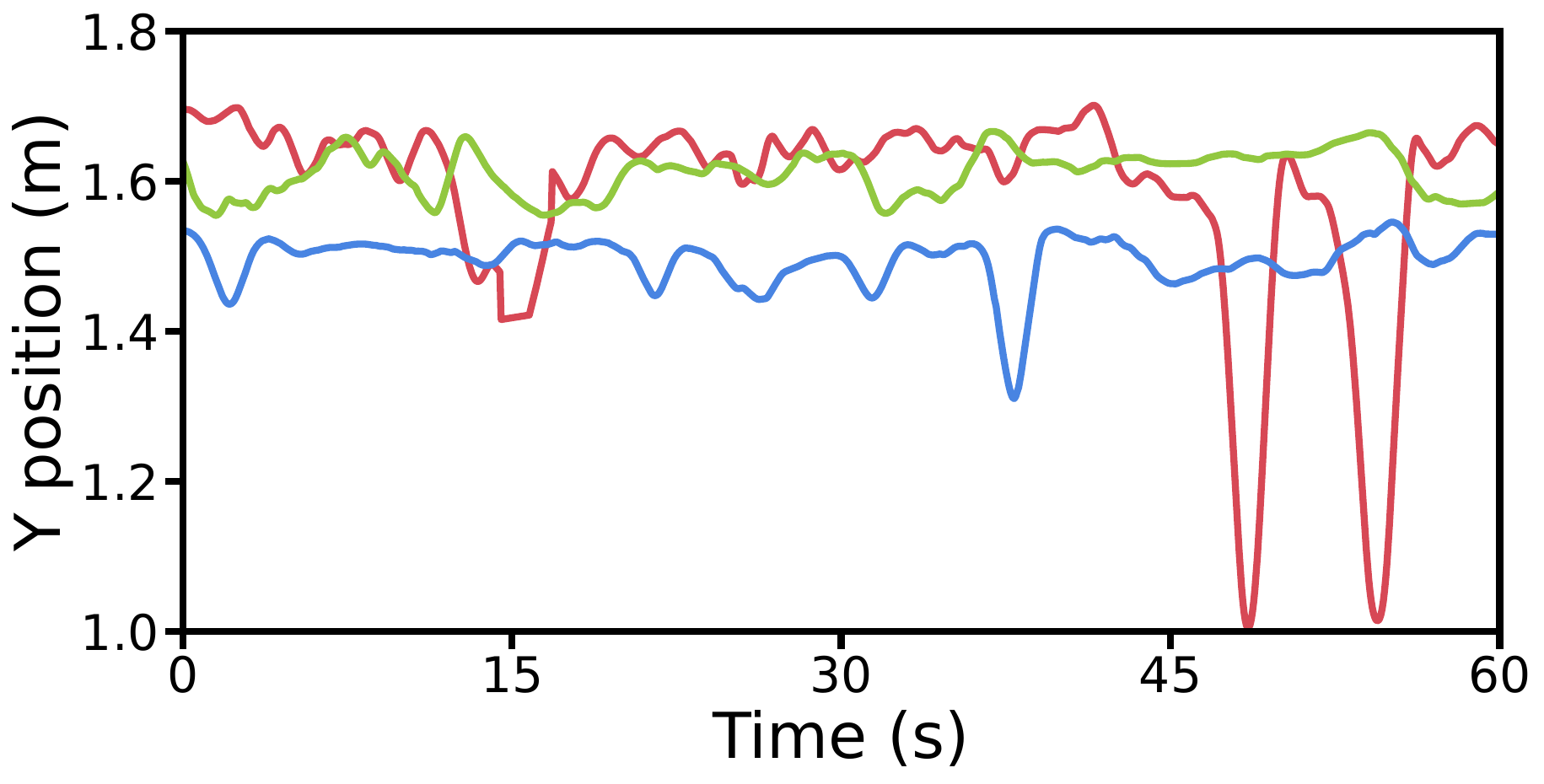} 
			\caption{nyc $\mathrm{Y}$-axis}
			\label{fig:scene2_y}
		\end{subfigure}
		\caption{User movement trajectories for an outdoor and an indoor scene, with three users per scene plotted with different colors. Top row: horizontal user movements (on the $\mathrm{XZ}$ plane) within the \threemeterarea area; Bottom row: corresponding changes of the headset height (along the $\mathrm{Y}$-axis) over time.}
		\label{fig:user_trajectory}
	\end{figure}

	Figure \ref{fig:user_trajectory} shows two examples of the participants' tracked position in the \threemeterarea physical stage area. 
	For each figure, we plot the traces of three users' movements in the 60-second period, in different colors. 
	Since movement along the vertical Y-axis (aligned with gravity) is much less pronounced than movement along the horizontal XZ-plane, we plot it on a separate graph for clarity.
	The average distances walked by the participants in the 60-second viewing sessions are reported in Table \ref{tbl:summary_metrics}.

	As the GPU and HMD connection methods are different at the RU and NTHU sites, the observed frame rates during trace collection also differ.
	As shown in Table \ref{tbl:summary_metrics}, the RU site, with its more powerful GPU for rendering, averaged approximately 57 frames per second (fps), compared to approximately 40 fps at the NTHU site.
	The average frame rates for each scene are provided in Table \ref{tbl:fps_results}.

	\section{Dataset Use Cases}
	
	The \name\ dataset, with its detailed 6-DoF navigation traces including head pose and eye gaze information, offers valuable opportunities for research in immersive computing systems.
	We describe example use cases in this section. 
	
	\vspace{0.3em}
	\noindent\textbf{6-DoF Viewport Prediction and Streaming Optimization.}
	The \name\ dataset fills the gap of 6-DoF user navigation traces of reconstructed real-world scenes that existing datasets ~\cite{KhanChakareski2020NJIT6DOF,subramanyam2020user,chakareski20206dof,Chen22VRViewportPose} lack.
	The fine-grained head pose and eye gaze of reconstructed real-world scenes in VR can be used for developing 6-DoF viewport prediction algorithms. 
	Such prediction can inform the design of adaptive media streaming algorithms to fetch only the content needed for rendering the user's viewport without wasting bandwidth on unviewed portion of the representation, e.g.,~\cite{gul2020kalman,zhu2025sgss,tsai2025l3gs}.

	\begin{table}[!t]
		\centering
		\caption{Statistics of Navigation Traces Per-session (60 Seconds) at Two Collection Sites}
		\vspace{-1em}
		\label{tbl:summary_metrics}
		\begin{tabular}{lcc}
			\toprule
			\textbf{Metric} & \textbf{RU Site} & \textbf{NTHU Site} \\
			\midrule
			Avg. \# of recorded frames & 3,420  & 2,396 \\
			Frames per second (fps) & 57.04  & 39.95 \\
			Total distance traveled (meters) & 17.27 & 13.62 \\
			\bottomrule
		\end{tabular}
	\end{table}
	
	\begin{table}[!t]
		\centering
		\caption{Average Frame Rates for Each Scene at Two Sites}
		\vspace{-1em}
		\label{tbl:fps_results}
		\resizebox{0.48\textwidth}{!}{%
			\begin{tabular}{@{}lcccccc@{}}
				\toprule
				\textbf{Site} & \textbf{alameda} & \textbf{berlin} & \textbf{bicycle} & \textbf{drjohnson} & \textbf{london} & \textbf{nyc} \\ \midrule
				\textbf{RU} & 43.68 & 55.33 & 42.12 & 37.47 & 65.32 & 44.18 \\
				\textbf{NTHU} & 36.68 & 40.62 & 33.15 & 33.81 & 45.86 & 37.66 \\ \bottomrule
				\toprule
				\textbf{Site} & \textbf{playroom} & \textbf{room} & \textbf{stump} & \textbf{train} & \textbf{treehill} & \textbf{truck} \\ \midrule
				\textbf{RU} & 47.53 & 71.37 & 68.80 & 71.30 & 70.91 & 69.75 \\
				\textbf{NTHU} & 36.67 & 41.80 & 44.43 & 43.65 & 41.63 & 43.17 \\ \bottomrule
			\end{tabular}
		}
	\end{table}

	\vspace{0.3em}
	\noindent\textbf{3D Saliency and Saliency in VR.}
	The rich per-frame eye gaze information in the \name\ dataset also offers opportunities for 3D saliency research~\cite{sitzmann2018saliency,wang2024saliency3d,ennadifi2023enhancing}. 
	This detailed fixation data can be aggregated across participants to create the groundtruth 3D saliency maps of the 3DGS scenes. 
	These maps can then be used for training 3D saliency models that better predict where the users will look in reconstructed real-world scenes. 
	Furthermore, since the reconstructed 3DGS scene may contain imperfections such as under-constructed areas and other visual artifacts, our dataset also enables studies into how these imperfections influence user gaze and navigation behavior. 
	
	\vspace{0.3em}
	\noindent\textbf{Foveated Rendering Optimization.}
	Foveated rendering is an important technique in VR designed to reduce the rendering computation demand and improve the frame rates~\cite{patney2016towards,krajancich2023towards,singh2023power}.
	Given that the visual acuity of human vision decreased sharply away from the foveal center, foveated rendering works by reducing the shading rates in the peripheral (non-foveal) region of the user's view. 
	This can achieve significant performance gain while with minimal impact on visual quality. 
	Existing works have explored applying foveated rendering for 3DGS rendering, e.g., MetaSapiens~\cite{lin2025metasapiens} and VR-Splatting~\cite{franke2025vr}. 
	Our \name\ dataset features per-frame eye gaze traces collected during free world standing 6-DoF navigation of 3DGS scenes, facilitating evaluation and optimization of real-world performance of these foveated rendering techniques.

	\section{Conclusion}
	In this paper, we present \name, the first publicly available 6-DoF navigation dataset built on photorealistic 3DGS reconstructions of real-world scenes, together with an open-source record-and-replay software fork of the SIBR viewer. 
	\name\  focuses on capturing detailed and realistic behavior.
	It does so by collecting user traces in scenes of real-world environments, carefully calibrated (tilt, scale, and starting viewpoint) to maintain realism.
	Our dataset captures stereoscopic pose, field-of-view, and eye-tracking traces.
	In addition, multi-site data collection (46 participants across two institutions) ensure a diverse population of participants.
	The realism, detail, and diversity of \name\ fills a critical gap for immersive media researchers,
	allowing user-centric evaluation in core areas such as streaming, rendering, and compression.

	\bibliographystyle{ACM-Reference-Format}
	\bibliography{ref}
	
\end{document}